\documentclass{pnastwo}

\usepackage{graphicx}
\usepackage{float}

\usepackage{amssymb,amsfonts,amsmath,color}

 \let\b=\beta

  \let\f=\varphi \let\ph=\varphi

\def\to{\rightarrow}

\newcommand{\beq}{\begin{equation}} \newcommand{\eeq}{\end{equation}}

\renewcommand{\phi}{\varphi}

\newcommand{\av}[1]{{\left\langle {#1} \right\rangle}}
\renewcommand{\vec}[1]{\boldsymbol{#1}}
\def\vr{{\boldsymbol r}}

\def\tw{t_{\mathrm{w}}}
\newcommand{\V}[1]{\boldsymbol{#1}}
\def\phid{\varphi_{\rm d}}
\def\phij{\varphi_{\rm J}}
\def\phiG{\varphi_{\rm G}}
\def\phiin{\varphi_{\rm g}}

\contributor{Submitted to Proceedings
of the National Academy of Sciences of the United States of America}
\url{www.pnas.org/cgi/doi/10.1073/pnas.XXXXXXXXXX}
\copyrightyear{2014}
\issuedate{Issue Date}
\volume{Volume}
\issuenumber{Issue Number}
\begin{document}

\title{Growing timescales and 
lengthscales characterizing vibrations of amorphous solids} 
 
\author{Ludovic Berthier\affil{1}{Universit\'e de Montpellier, Montpellier, France},
Patrick Charbonneau\affil{2}{Department of Chemistry, Duke University, Durham,North Carolina 27708, USA}
\affil{3}{Department of Physics, Duke University, Durham, North Carolina 27708, USA},
Yuliang Jin 
\thanks{To whom correspondence should be addressed. E-mail: jinyuliang@gmail.com}
\affil{2}{Department of Chemistry, Duke University, Durham, North Carolina 27708, USA}
\affil{4}{Dipartimento di Fisica, Sapienza Universit\'a di Roma, INFN, Sezione di Roma I, IPFC -- CNR, Piazzale Aldo Moro 2, I-00185 Roma, Italy}
\affil{5}{Laboratoire de Physique Th\'eorique, ENS \& PSL University, UPMC \& Sorbonne Universit\'es, 
UMR 8549 CNRS, 75005 Paris, France},
Giorgio Parisi
\affil{4}{Dipartimento di Fisica, Sapienza Universit\'a di Roma, INFN, Sezione di Roma I, IPFC -- CNR,  Piazzale Aldo Moro 2, I-00185 Roma, Italy},
Beatriz Seoane 
\thanks{To whom correspondence should be addressed. E-mail:beatriz.seoane.bartolome@lpt.ens.fr}
\affil{5}{Laboratoire de Physique Th\'eorique, ENS \& PSL University, UPMC \& Sorbonne Universit\'es, 
  UMR 8549 CNRS, 75005 Paris, France},
\affil{6}{Instituto de Biocomputaci\'on y F\'{\i}sica de Sistemas Complejos (BIFI), 50009 Zaragoza, Spain}
\and
Francesco Zamponi
\affil{5}{Laboratoire de Physique Th\'eorique, ENS \& PSL University, UPMC \& Sorbonne Universit\'es, 
UMR 8549 CNRS, 75005 Paris, France},
}

\contributor{Submitted to Proceedings of the National Academy of Sciences
of the United States of America}

\maketitle

\begin{article}

\begin{abstract}
Low-temperature properties of crystalline solids 
can be understood using harmonic perturbations around a perfect
lattice, as in Debye's theory. 
Low-temperature properties of amorphous solids, however, strongly depart from 
such descriptions, displaying enhanced transport,
activated slow dynamics across energy barriers,
excess vibrational modes with respect to Debye's 
theory (i.e., a Boson Peak),
and complex irreversible responses to small 
mechanical deformations.
These experimental observations indirectly suggest that the dynamics of amorphous 
solids becomes anomalous at low temperatures.
Here, we present direct numerical evidence that vibrations
change nature at a well-defined location deep inside the glass phase of a simple glass former. We provide a real-space 
description of this transition and of the rapidly growing time 
and length scales that accompany it. 
Our results provide the seed for a universal understanding of low-temperature glass anomalies
within the theoretical framework of the recently 
discovered Gardner phase transition.
\end{abstract}


\noindent \noindent\rule{8cm}{0.4pt}
\\
{\bf Significance}
Amorphous solids constitute most of 
solid matter but remain poorly understood.
The recent solution of the mean-field hard sphere glass former provides,
 however, deep insights into their material properties. In
particular, this solution predicts a Gardner transition below which
the energy landscape of glasses becomes fractal and the solid is
marginally stable. Here we provide the first direct evidence for the
relevance of a Gardner transition in physical systems. This result thus
opens the way towards a unified understanding of the
low-temperature anomalies of amorphous solids.
\noindent\rule{8cm}{0.4pt}

\paragraph*{Introduction -}
Understanding the
nature of the glass transition, which describes the gradual transformation 
of a viscous liquid into an amorphous solid, remains 
an open challenge in condensed matter physics~\cite{BB11,Ca09}. As a result,   
the glass phase itself is not well 
understood either. The main challenge is to connect the localised, 
or `caged', dynamics that characterizes the glass transition 
to the low-temperature anomalies
that distinguish amorphous
solids from their crystalline counterparts~\cite{Phillips1987,Goldstein2010,MS86,HKEP11,MW15}. 
Recent theoretical advances, building on the random 
first-order transition approach~\cite{WL12}, have led to an 
exact mathematical description of both the glass transition and the 
amorphous phases of hard spheres in the mean-field limit 
of infinite-dimensional space~\cite{CKPUZ14}. 
A surprising outcome has been the discovery of a novel phase 
transition inside the amorphous phase, separating the 
localised states produced 
at the glass transition from their inherent structures.
This Gardner transition~\cite{Ga85}, which
marks the emergence of a fractal hierarchy of marginally stable glass states,
can be viewed as a glass transition deep within a glass, 
at which vibrational motion dramatically slows down 
and becomes spatially
correlated~\cite{CJPRSZ2015PRE}. Although these theoretical findings 
promise to explain and unify the emergence of low-temperature  
anomalies in amorphous solids, the 
gap remains wide between mean-field calculations~\cite{CKPUZ14,CJPRSZ2015PRE} 
and experimental work.
Here, we provide direct numerical evidence 
that vibrational motion in a simple three-dimensional  
glass-former becomes anomalous at a well-defined 
location inside the glass phase. In particular, we 
report the rapid growth of a relaxation
time related to cooperative vibrations, a non-trivial change in the probability
distribution function of a global order parameter,
and the rapid growth of a correlation length.
We also relate these findings to observed anomalies in
low-temperature laboratory glasses.
These results provide key support for a universal understanding of the anomalies of glassy materials, 
as resulting from the diverging length and time scales associated with the criticality
of the Gardner transition.

\begin{figure}
\centerline{\includegraphics[width=1\columnwidth]{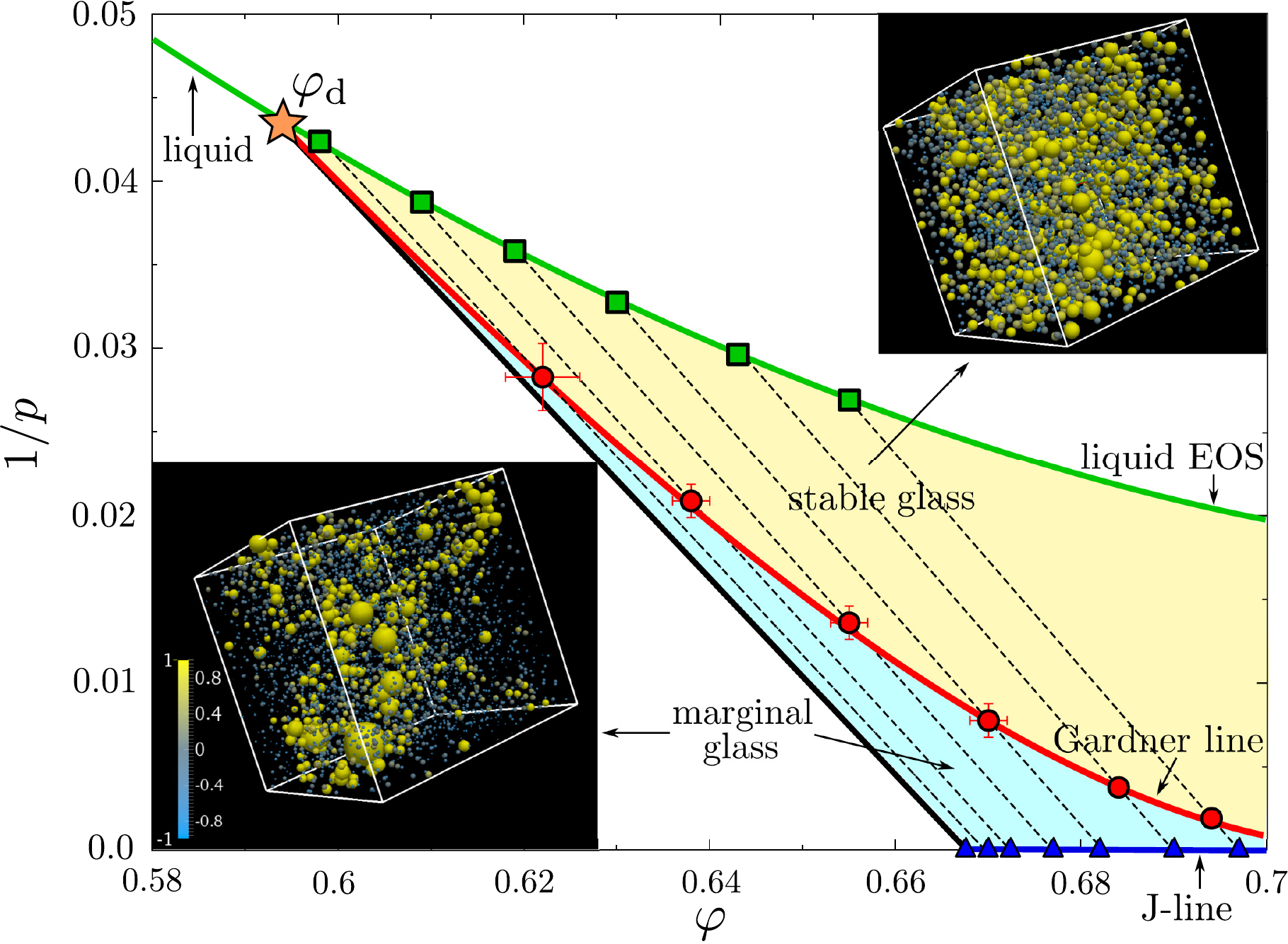} }
\caption{{\bf Two glass phases.} Inverse reduced pressure--packing fraction 
($1/p$ v $\varphi$) phase diagram for polydisperse hard
spheres. The equilibrium simulation results at $\varphi_{\rm g}$ 
(green squares) are fitted to the liquid EOS (Eq.~\eqref{eq:liquid_EOS}, green line). 
The dynamical crossover, $\varphi_\mathrm{d}$, 
is obtained from the liquid dynamics. 
Compression annealing from $\varphi_{\rm g}$ 
up to jamming (blue triangles) follows a glass
EOS (fit to Eq.~\eqref{eq:glass_EOS}, dashed lines).  
At $\varphi_{\mathrm{G}}$ (red circles and line) with a finite $p$, stable glass states transform into marginally 
stable glasses. 
Snapshots illustrate spatial heterogeneity
above and below $\varphi_{\mathrm{G}}$, 
with sphere diameters proportional to the linear cage 
size and colors encoding the 
relative cage size, $u_i$ (see text).
\label{fig:PD}}
\end{figure}

\begin{figure}
\centerline{\includegraphics[width=1\columnwidth]{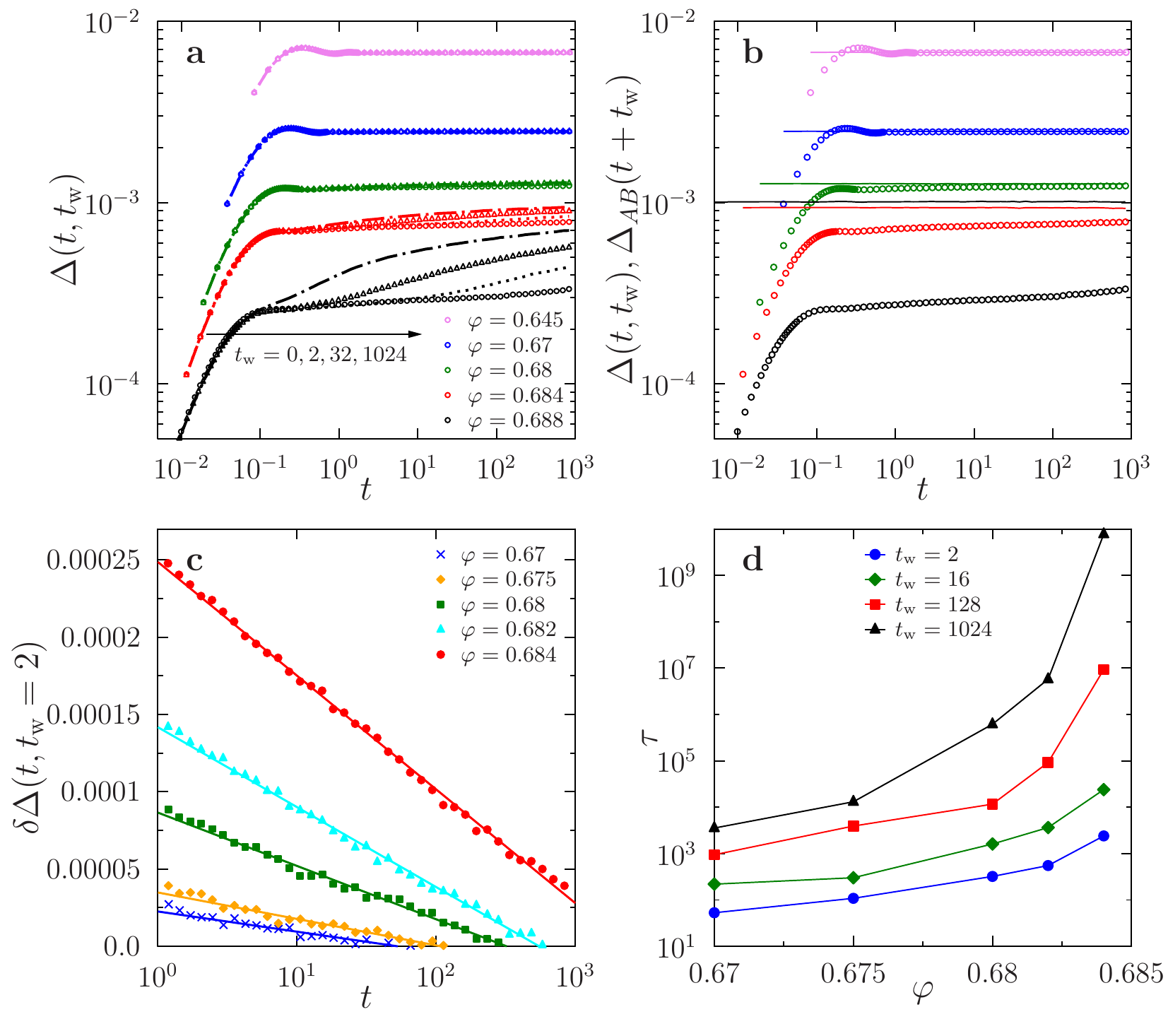} }
\caption{{\bf Emergence of slow vibrational dynamics.} (a) Time evolution of
  $\Delta(t,\tw)$ for several $\tw$ and $\varphi$, following 
  compression from $\varphi_{\rm g}=0.643$. For
  $\varphi\gtrsim\varphi_{\mathrm{G}}=0.684$, $\Delta(t,\tw)$
  displays strong aging. (b) Comparison between
  $\Delta(t,\tw)$ (points) and $\Delta_{AB}(t+\tw)$ (lines) for the longest waiting
  time $\tw=1024$. For $\varphi<\varphi_{\mathrm{G}}$ both observables
  converge to the same value within the time window considered, but not
  for $\varphi>\varphi_{\mathrm{G}}$. (c) The time evolution of $\delta \Delta(t, \tw)$
  at
  $\tw=2$ displays a logarithmic tail, which
  provides a characteristic relaxation time $\tau$.
  (d) As 
 $\varphi$ approaches $\phi_{\rm G}$, $\tau$ grows rapidly for any $\tw$.
\label{fig:dynamics}}
\end{figure}

\begin{figure}
\centering\includegraphics[width=1\columnwidth]{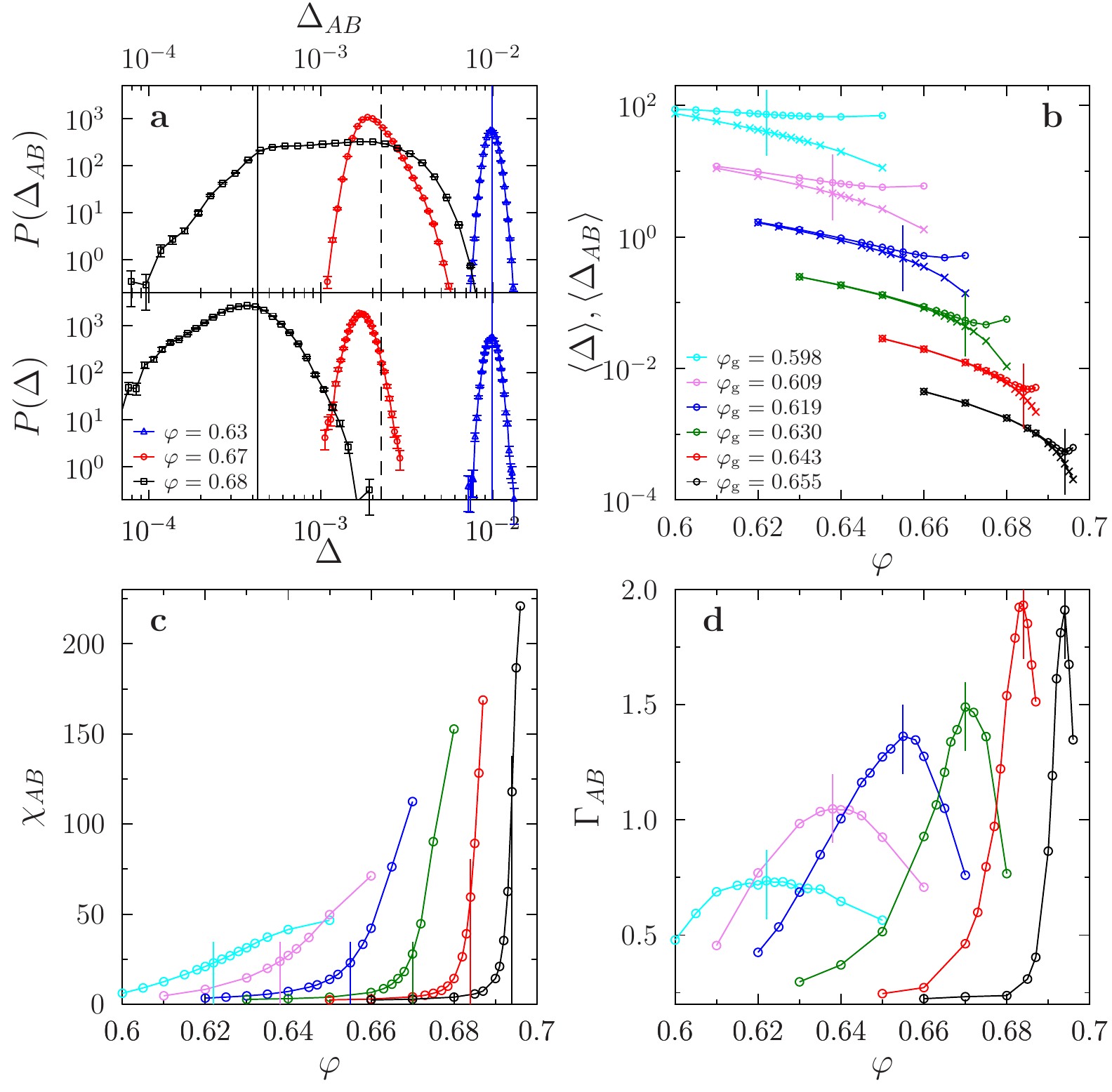} 
\caption{{\bf Global fluctuations of the order parameter.}
(a) Probability distribution functions for
  $\Delta_{AB}$ and $\Delta$ above, at, and below
  the Gardner crossover, $\varphi_\mathrm{G}=0.670(2)$ for
  $\varphi_{\rm g}=0.630$. Vertical lines mark 
  $\langle \Delta \rangle$ (solid) and $\langle \Delta_{AB} 
\rangle$ (dashed), which also represent the peak positions.
  (b) Comparing $\langle \Delta \rangle $ and
  $\langle \Delta_{AB} \rangle $ shows that the average values separate for
  $\varphi\gtrsim\varphi_\mathrm{G}$  (Data are multiplied by  $5^k$, where $k=0, 1, \ldots, 5$ for $\varphi = 0.655, 0.643, \ldots, 0.598$ respectively). Around $\varphi_{\mathrm{G}}$, 
(c) the global susceptibility $\chi_{AB}$ grows very rapidly, 
 and (d) the skewness 
  $\Gamma_{AB}$ peaks. Numerical estimates for $\varphi_\mathrm{G}$ are 
indicated by vertical segments.
\label{fig:statics}}
\end{figure}

\begin{figure}
\centering{\includegraphics[width=1\columnwidth]{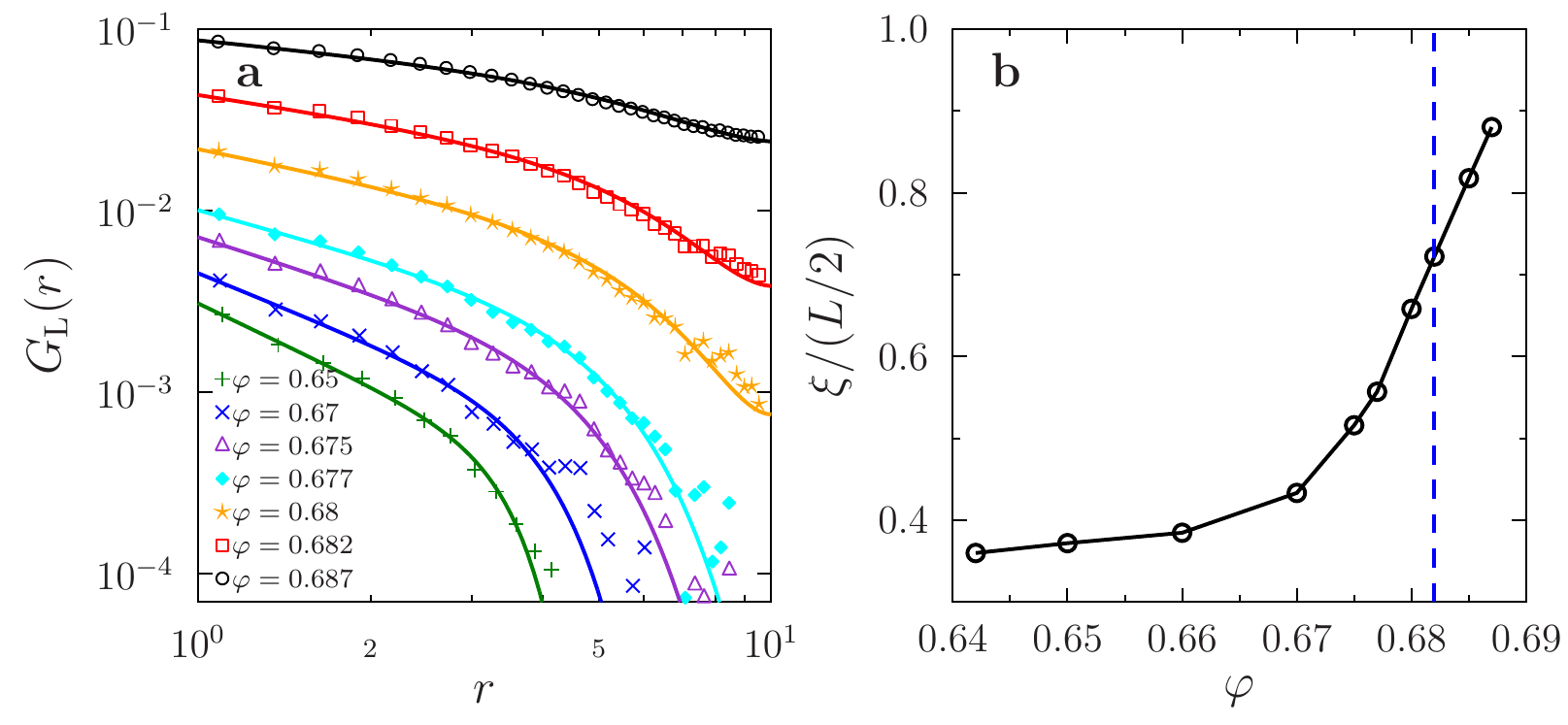}}
\caption{{\bf Growing correlation length.} (a) Spatial
  correlator $G_\mathrm{L}(r)$ (Eq.~\eqref{eq:G})
  for different $\varphi$ annealed from $\varphi_{\rm g}=0.640$, 
with $N=8000$ (larger systems are used here to significantly measure 
the growth of $\xi$). 
  (b) Fitting $G_\mathrm{L}(r)$ to
  Eq.~\eqref{eq:Glongr}   (lines in (a)) provides the correlation length
  $\xi$ , which grows with 
$\varphi$ and becomes comparable to the  linear system size upon approaching $\varphi_{\rm G}=0.682(2)$ (dashed line).
\label{fig:heterogeneity}}
\end{figure}

\paragraph*{Preparation of glass states --}
Experimentally, glasses are obtained by
a slow thermal or compression annealing, the rate of which
determines the location of the glass transition~\cite{BB11,Ca09}. We find that
a detailed numerical analysis of the Gardner transition requires
the preparation of extremely well-relaxed glasses
(corresponding to structural relaxation timescales challenging to simulate)
in order to study vibrational motion inside 
the glass without interference from 
particle diffusion. We thus combine a 
very simple glass-forming model -- 
a polydisperse mixture of hard spheres -- to
an efficient Monte-Carlo scheme 
to obtain equilibrium configurations at unprecedentedly high densities, i.e., deep in 
the supercooled regime. 
The optimized swap Monte-Carlo 
algorithm~\cite{grigera2001fast}, which combines 
standard local Monte-Carlo moves with attempts at exchanging pairs of 
particle diameters, indeed enhances thermalization 
by several orders of magnitude.
Configurations contain either $N=1000 $ or $N=8000$ (results 
in Figs.~\ref{fig:PD}-\ref{fig:statics} are for $N=1000$, and
for $N=8000$ in Fig.~\ref{fig:heterogeneity}) hard spheres with equal unit mass $m$ and diameters 
independently drawn from a probability distribution  $P_{\sigma}(\sigma) \sim 
\sigma^{-3}$, for $\sigma_{\rm min} \leq \sigma  \leq \sigma_{\rm min} /0.45$. 
 We similarly study a two-dimensional bidisperse model glass former and report the main results in the Appendix.

We mimic slow annealing 
in two steps (Fig.~\ref{fig:PD}).
First, we produce equilibrated liquid configurations
at various densities  $\f_{\rm g}$
using our efficient
simulation scheme, concurrently obtaining the
liquid equation of state (EOS). 
The liquid EOS for the reduced pressure $p=\beta P/\rho$, 
where $\rho$ is
the number density,  $\beta$ is the inverse temperature, and $P$ is the system pressure, is described by
\beq
p_{\rm liquid} (\varphi) = 1+f(\varphi) [p_{\rm CS}(\varphi)-1],
\label{eq:liquid_EOS}
\eeq
with $p_{\rm CS}(\varphi)$ from Ref.~\cite{B70}:
\beq
p_{\rm CS}(\varphi) = \frac{1}{1-\varphi} + \frac{3 s_1 s_2}{s_3} 
\frac{\varphi}{(1-\varphi)^2} + \frac{s_2^3}{s_3^2} \frac{(3-\varphi)
\varphi^2}{(1-\varphi)^3},
\label{eq:poly_EOS}
\eeq
where
 $s_k$ is the $k$-th moment of $P_{\sigma}(\sigma)$, and $f(\varphi)=0.005-
\tanh[14(\varphi-0.79)]$ are fitted quantities.
The structure of the equilibrium configurations generated by the swap algorithm has been carefully analyzed. Unlike for other glass formers~\cite{RT2015,FS2015}, no signs of 
 orientational or crystalline order were observed~\cite{BCNO2015,YBCT2015}. Following the strategy of Ref.~\cite{Charbonneau2014}, we also obtain the 
 dynamical crossover $\varphi_{\rm d} = 0.594(1)$ (see the Appendix). 
 We have not analyzed the compression of 
equilibrium configurations with $\varphi_{\rm g} < \varphi_{\rm d}$,
as done in earlier studies~\cite{CBS2010, OKIM2012}, 
because structural relaxation is not well decoupled from vibrational
dynamics, although the obtained jammed states 
should have equivalent properties.

Second, we use these liquid configurations as starting points for
standard molecular dynamics simulations during which
the system is compressed out of equilibrium up to various target $\f > \f_{\rm g}$~\cite{SDST06}.
Annealing 
is achieved by growing spheres following the
Lubachevsky-Stillinger (LS) algorithm~\cite{SDST06} at a constant 
growth rate $\gamma_{\rm g} =10^{-3}$ (see the Appendix for a discussion on the $\gamma_{\rm g}$-dependence).
The average particle diameter, $\bar\sigma$, serves as unit length, and
the simulation time is expressed in units of $\sqrt{\beta m \bar{\sigma}^2}$, with unit inverse temperature $\beta$  and unit mass $m$.
In order to obtain thermal and disorder averaging, this
procedure is repeated over $N_\mathrm{s}$ samples ($N_\mathrm{s}\approx150$ for
$N=1000$ and $N_\mathrm{s}=50$ for $N=8000$), 
each with different initial equilibrium configurations at $\varphi_{\rm g}$, and over $N_\mathrm{th} = 64 - 19440$ independent thermal (quench)
histories for each sample. Quantities reported here are averaged
over $ N_\mathrm{s} \times N_\mathrm{th}$ quench histories, unless
otherwise specified. 
The non-equilibrium glass EOSs
associated with this compression (dashed lines) terminate 
(at infinite pressure) at inherent structures that correspond, for hard spheres, 
to jammed configurations (blue triangles).
In order to capture the glass
EOSs, we use a free volume scaling around the corresponding jamming point $\varphi_{\rm J}$,
\beq\label{eq:glass_EOS}
p_{\rm glass} (\varphi ) = \frac{C}{\varphi_{\rm J} - \varphi},
\eeq
where the constant $C$ weakly depends on $\varphi_{\rm g}$.

Our numerical protocol is analogous to varying
the cooling rate -- and thus the glass transition temperature
-- of thermal glasses, 
and then further annealing the resulting amorphous solid. 
Each value of $\f_{\rm g}$ indeed selects a different 
glass, ranging from the onset of 
sluggish liquid dynamics
around the mode-coupling theory dynamical crossover~\cite{Ca09,BB11}, $\f_{\rm d}$, 
to the very dense liquid regime where
diffusion and vibrations ($\b$-relaxation processes) are fully separated~\cite{Ca09}. 
For sufficiently large $\f_{\rm g}$, 
we thus obtain unimpeded access to the only 
remaining glass dynamics, i.e., $\b$-relaxation 
processes~\cite{Goldstein2010}.

\paragraph*{Growing timescales --}
A central observable to characterize glass dynamics 
is the mean-squared displacement (MSD) of particles from position $\vr_i(\tw)$, 
\beq
\Delta (t,\tw) =
\frac{1}{N} \sum _{i=1}^{N} \av{|\vr_i(t + \tw) - \vr_i(\tw)|^2}, 
\label{eq:MSD}
\eeq
averaged over both thermal fluctuations and disorder, 
where time $t$ starts after waiting time $\tw$ when compression has reached the target $\f$.
The MSD plateau height at long times quantifies the average cage size (see the Appendix). 
 Because some of the smaller particles manage to 
leave their cages, the sum in Eq.~\eqref{eq:MSD}
is here restricted to the larger half of the particle 
size distribution (see the Appendix).
When $\varphi$
is not too large, 
$\varphi \gtrsim \varphi_{\rm g}$, the plateau 
emerges quickly, 
as suggested by the traditional 
view of caging in glasses (Fig.~\ref{fig:dynamics}a). 
When the glass is compressed beyond a
certain $\varphi_\mathrm{G}$, however, $\Delta
(t,\tw)$ displays both a strong dependence on the waiting time $\tw$, i.e., aging, and a slow dynamics, as 
captured by the emergence of two plateaux.

These effects suggest a complex vibrational 
dynamics. Aging, in particular, provides a striking signature 
of a growing timescale associated with vibrations,
revealing the existence of a ``glass transition'' 
deep within the glass phase.

To determine the timescale associated with this slowdown, 
we estimate the distance between independent pairs 
of configurations
by first compressing two independent copies, $A$ and $B$, 
from the same initial state at $\phi_{\rm g}$ to the target $\varphi$,
and then measuring their relative distance 
\beq
\Delta_{AB} (t) = \frac{1}{N}
\sum _{i=1}^{N} \av{|\vr_i^A(t) - \vr_i^B(t)|^2},
\eeq 
so that 
$\Delta_{AB} (t \to \infty) \simeq 
\Delta (t \to \infty, \tw \to \infty)$, as shown in 
Fig.~\ref{fig:dynamics}b.
The two copies share the same positions of particles 
at $\phi_{\rm g}$, but are assigned different initial 
velocities drawn from the Maxwell--Boltzmann distribution.
The time evolution of the difference 
$\delta \Delta(t, \tw) = \Delta_{AB}(\tw+t) - \Delta(t, \tw)$
indicates that whereas the amplitude of particle motion naturally becomes 
smaller as $\phi$ increases, the corresponding dynamics 
becomes slower (Fig.~\ref{fig:dynamics}c). In other words, as $\phi$ grows particles take longer to 
explore a smaller region of space. In a crystal, by contrast, $\delta \Delta(t,\tw)$ 
decays faster under similar circumstances.
A relaxation timescale, $\tau$, can be extracted from the decay of $\delta \Delta(t, \tw)$ 
at large $t$, whose logarithmic
form,
$\delta \Delta(t, \tw)\sim  1 - \ln t /\ln \tau$,
is characteristic of the glassiness of vibrations. As $\phi \to \phi_{\rm G}$,
we find that $\tau$ dramatically increases (Fig.~\ref{fig:dynamics}d), which 
provides direct evidence of a marked crossover 
characterizing the evolution of the glass upon compression.

\paragraph*{Global fluctuations of the order parameter --} 
This sharp dynamical crossover corresponds to  
a loss of ergodicity inside the glass, i.e., time and ensemble averages yield different
results. To better characterize this crossover, we define a 
timescale $\tau_{\rm cage}$ for the onset of caging ($\tau_{\rm cage}
\approx {\cal O}(1)$, see the Appendix), and the corresponding 
order parameters $\Delta_{AB} \equiv \Delta_{AB} (\tau_{\rm cage})$
and $\Delta \equiv \Delta(\tau_{\rm cage},\tw =0)$.

The evolution of the probability distribution functions, $P(\Delta_{AB})$ and 
$P(\Delta)$, as well as their first moments, $\langle \Delta_{AB} \rangle$ 
and $\langle \Delta \rangle$, are presented in Figs.~\ref{fig:statics}a,b
for a range of densities across $\phi_{\rm G}$. For $\phi < \phi_{\rm G}$,
dynamics is fast, $\langle \Delta_{AB} \rangle$ 
and $\langle \Delta \rangle$ coincide, and $P(\Delta_{AB})$ and 
$P(\Delta)$ are narrow and Gaussian-like. For $\phi > \phi_{\rm G}$, however, 
the MSD does not converge to its long-time limit,  
 $\langle \Delta \rangle < \langle \Delta_{AB} \rangle$,
which indicates that configuration space explored by vibrational 
motion is now broken into mutually inaccessible regions.  Interestingly, the slight increase of $\langle \Delta_{AB} \rangle$ with $\varphi$ in this regime (Fig.~\ref{fig:statics}b) suggests that states are then pushed further apart in phase space, which is consistent with theoretical predictions~\cite{CJPRSZ2015PRE}.
When compressing a system across $\phi_{\rm G}$, 
its dynamics  explores only a restricted part of phase space. 
As a result, $\Delta_{AB}$ displays
pronounced,  non-Gaussian 
fluctuations
(Fig.~\ref{fig:statics}a). Repeated compressions 
from a same initial state at $\phi_{\rm g}$ may end up
in distinct states, which explains why $\Delta_{AB}$
is typically much larger and more broadly fluctuating than $\Delta$ 
(Fig.~\ref{fig:statics}a).  These results are essentially consistent with theoretical predictions~\cite{CKPUZ14,CJPRSZ2015PRE}, which suggest that  for $\varphi > \varphi_{\rm G}$, $P(\Delta_{AB})$ should separate into two peaks connected by a wide continuous band with the left-hand peak continuing the single peak of $P(\Delta)$.
The very broad 
 distribution of $\Delta_{AB}$
further suggests that spatial correlations 
develop as $\phi \to \phi_{\rm G}$, yielding strongly 
correlated states at larger densities.

To quantify these fluctuations we measure the 
variance $\chi_{AB}$ and skewness $\Gamma_{AB}$ (see the Appendix and Ref.~\cite{CJPRSZ2015PRE} ) of 
$P(\Delta_{AB})$ (Figs.~\ref{fig:statics}c,d). 
The global susceptibility $\chi_{AB}$
is very small for $\phi < \phi_{\rm G}$ and grows 
rapidly as $\phi_{\rm G}$ is approached, increasing by 
about two decades for the largest 
$\varphi_{\rm g}$ considered (Fig.~\ref{fig:statics}c). 
While $\chi_{AB}$ quantifies the increasing 
width of the distributions, $\Gamma_{AB}$ reveals a change 
in their shapes. For each $\phi_{\rm g}$ we find 
that $\Gamma_{AB}$ is small on both sides 
of $\phi_{\rm G}$ with a pronounced maximum at $\phi = \phi_{\rm G}$
(Fig.~\ref{fig:statics}d). This reflects the roughly symmetric shape of
$P(\Delta_{AB})$ around $\langle \Delta_{AB} \rangle$
on both sides of $\phi_{\rm G}$ and the development of an asymmetric tail for 
large $\Delta_{AB}$ around the crossover, a 
known signature of sample-to-sample fluctuations
in spin glasses~\cite{PR12} and mean-field 
glass models~\cite{CJPRSZ2015PRE}.
Note that because the skewness maximum gives the clearest 
numerical estimate of $\varphi_{\mathrm{G}}$, 
we use it to determine the values reported in 
Fig.~\ref{fig:PD}.

\paragraph*{Growing correlation length --} 
The rapid growth of $\chi_{AB}$ in the vicinity
of $\phi_{\rm G}$ suggests the concomitant growth of a spatial correlation
length, $\xi$. Its measurement requires 
spatial resolution of the fluctuations of $\Delta_{AB}$, hence
for each particle $i$ we define 
$u_i=\frac{ |\vec{r}_i^A -  \vec{r}_i^B|^2}{\av{{\Delta}_{AB}}}-1$
to capture its contribution to deviations 
around the average $\langle \Delta_{AB} \rangle$. 
A first glimpse of these spatial fluctuations is offered 
by snapshots of the $u_i$ field 
(Fig.~\ref{fig:PD}), which appear featureless for $\varphi<\varphi_{\rm G}$,
but highly structured and spatially correlated for $\varphi\gtrsim\phi_{\rm G}$. 
More quantitatively, we define the spatial 
correlator,  
\beq G_{\rm L}(r) = \frac{\av{\sum_{\mu =
      1}^{3} \sum_{i\neq j}u_i
    u_j\delta\left(r-|\vec{r}_{i,\mu}^A-\vec{r}_{j,\mu}^A|\right)}}{\av{
\sum_{\mu
      = 1}^{3} \sum_{i\neq j}
    \delta\left(r-|\vec{r}_{i,\mu}^A-\vec{r}_{j,\mu}^A|\right)}},
\label{eq:G}
\eeq 
where $\vec{r}_{i,\mu}$ is the projection of the particle position along 
direction $\mu$. Even for the larger system size considered, 
measuring $G_{\rm L}(r)$ is challenging because spatial correlations 
quickly become long ranged as $\phi \to \phi_{\rm G}$ (see Fig.~\ref{fig:heterogeneity}a). 
Fitting the results to 
an empirical form that takes into account the periodic boundary conditions in a system of linear size $L$,
\begin{equation}\label{eq:Glongr}
G_{\rm L}(r) \sim \frac{1}{r^a}  \mathrm{e}^{-\left(\frac{r}{\xi}\right)^b}+ \frac{1}{(L-r)^a}
\mathrm{e}^{-\left(\frac{L-r}{\xi}\right)^b} \ ,
\end{equation}
where $a$ and $b$ are fitting parameters, 
nonetheless confirms 
that $\xi$ grows rapidly with $\varphi$
and becomes of the order of the simulation box at
$\varphi > \varphi_{\rm G}$ (Fig.~\ref{fig:heterogeneity}b). 
Note that although probed using a 
dynamical observable, the spatial correlations 
captured by $G_{\rm L}(r)$ are conceptually distinct from 
the dynamical heterogeneity observed in supercooled liquids~\cite{BBBCS11},
which is transient and disappears once the diffusive 
regime is reached.   

\paragraph*{Experimental consequences --}
The system analysed in this work is a canonical 
model for colloidal suspensions and granular media. 
Hence, experiments along the lines presented here
could be performed to investigate more closely 
vibrational dynamics in colloidal and granular glasses,
using a series of compressions to extract 
$\Delta$ and $\Delta_{AB}$.  Experiments are also possible in molecular and  
polymeric glasses, for which the natural control parameter is 
temperature $T$ instead of density. 
Let us therefore rephrase our findings from this viewpoint.
As the system is cooled, the supercooled liquid dynamics
is arrested at the laboratory glass transition temperature 
$T_{\rm g}$. 
As the resulting glass is further cooled  its phase space transforms, 
around a well-defined Gardner temperature  $T_{\rm G} < T_{\rm g}$, from a simple
state (akin to that of a crystal) into a more complex phase composed of a large number 
of 
glassy states 
(see the Appendix for a discussion of the phase diagram as a function of $T$).

Around $T_{\rm G}$, vibrational dynamics becomes
increasingly heterogeneous (Fig.~\ref{fig:PD}), slow 
(Fig.~\ref{fig:dynamics}),  
fluctuating from realization to realization (Fig.~\ref{fig:statics}), and spatially correlated
(Fig.~\ref{fig:heterogeneity}). 
The $\b$-relaxation dynamics inside the glass thus 
becomes highly cooperative~\cite{CKPS12,BKMPSR13} 
and ages~\cite{LN98b}.
The fragmentation of phase space below $T_{\rm G}$ 
also  gives rise to a complex response to mechanical perturbations
in the form of plastic irreversible events, in which the system jumps 
from one configuration to another~\cite{Goldstein2010,HKEP11,BW09b}. 
This expectation stems from the theoretical
prediction that the complex phase at $T < T_{\rm G}$  
is marginally stable~\cite{CKPUZ14}, which implies 
that glass states are connected by very low 
energy barriers, resulting in strong responses to 
weak perturbations~\cite{MW15}.  

A key prediction is that the aforementioned anomalies 
appear simultaneously around a $T_{\rm G}$
that is strongly dependent on the scale $T_{\rm g}$ selected by the glass 
preparation protocol. Annealed glasses with 
lower $T_{\rm g}$ are expected to present a sharper 
Gardner-like crossover, at an increasingly 
 lower
temperature. 
Numerically, we produced a 
substantial variation of $\f_{\rm g}$ by using an efficient 
Monte-Carlo algorithm to bypass the need for a broad range of 
compression rates.
In experiments a similar or even larger range of $T_{\rm g}$ can be explored~\cite{RSKQ94}, 
using poorly annealed glasses 
from hyperquenching~\cite{VBA01} and
ultrastable glasses from vapor deposition~\cite{ultra1,Singh2013,Hocky2014}. 
We expect ultrastable glasses, in particular, 
to display strongly 
 enhanced
glass anomalies, consistent with recent
experimental reports~\cite{PRRR14,LQMJH14,YTGER15}. Interestingly, a
Gardner-like regime may also underlie the anomalous 
aging recently observed in individual proteins~\cite{hu:2015}.

\paragraph*{Conclusion --} Since its prediction in the mean-field
limit, the Gardner transition has been regarded as a key ingredient 
to understand the physical properties of amorphous solids. 
Understanding the role of finite dimensional fluctuations is 
a difficult theoretical problem~\cite{UB14}.  
Our work shows that clear signs of an apparent critical behaviour
can be observed in three dimensions, at least in a finite-size 
system, which shows that the correlation length becomes 
at least comparable to the system size as $\varphi$ approaches 
$\varphi_{\rm G}$.  Although the fate of 
these findings in the thermodynamic limit remains an open question, the 
remarkably large signature of the effect strongly suggests that the 
Gardner phase transition paradigm 
is a promising theoretical framework for a universal understanding of the 
anomalies of solid amorphous materials, 
from granular materials to glasses, foams and proteins.

\begin{acknowledgments}

P.C. acknowledges support from the Alfred P. Sloan Foundation and 
NSF support No. NSF DMR-1055586. B.S. acknowledges the support 
by MINECO (Spain) through research contract No. FIS2012-35719-C02. 
This project has received funding from the European Union's Horizon 
2020 research and innovation programme under the Marie Sk{\l}odowska-Curie
grant agreement No. 654971 as well as from the 
European Research Council under the European Union's Seventh Framework 
Programme (FP7/2007-2013)/ERC grant agreement No. 306845. This work was granted access to the HPC resources of MesoPSL financed by the Region Ile de France and the project Equip@Meso (reference ANR-10-EQPX-29-01) of the programme Investissements d'Avenir supervised by the Agence Nationale pour la Recherche.
\end{acknowledgments}

\vspace{7cm}
\pagebreak

\appendix
\tableofcontents

\bigskip
All the results discussed in  this Appendix have been obtained using molecular dynamics (MD) simulations,
starting from the initial states produced using the swap algorithm as explained in the  main text.

\section{Dynamical crossover density}\label{sec:phid}

\begin{figure}[h]
\centerline{\hbox{\includegraphics [width = 1\columnwidth] {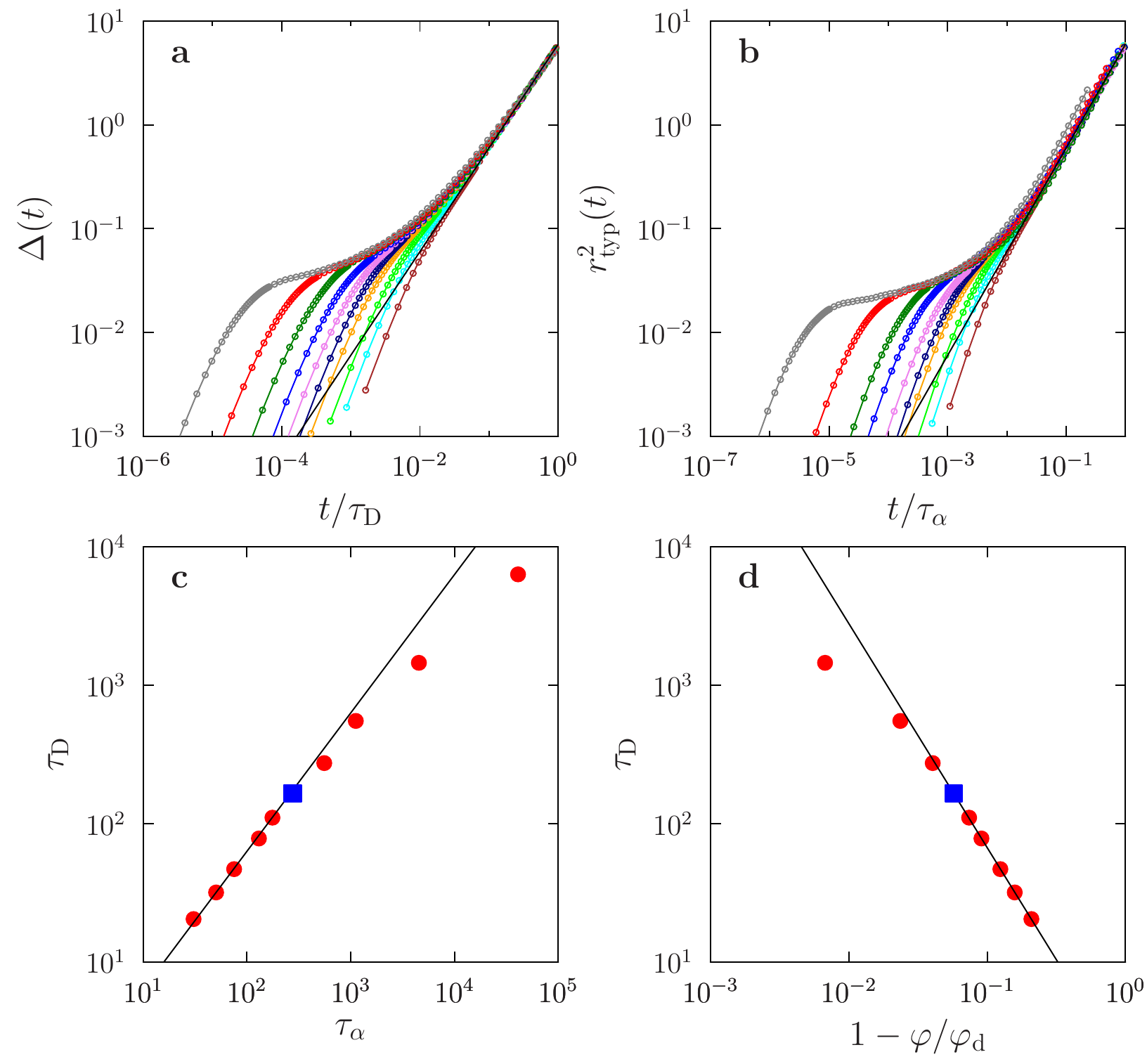}}}
\caption{Determination of 	$\varphi_{\rm d}$. Rescaled plots of the (a) MSD $\Delta(t)$ and (b) MSTD $r^2_{\rm typ}(t)$ at (from right to left) $\varphi=0.47, 0.50, 0.52, 0.54, 0.55, 0.56, 0.57, 0.58, 0.59, 0.60$. Solid black lines capture the long-time diffusive behavior, $\Delta(t) = 2d \bar{\sigma}^2 (t/\tau_D)$ and $r^2_{\rm typ}(t) = 2d \bar{\sigma}^2 (t/\tau_\alpha)$, respectively. (c) The SER (line) breaks down around $\varphi_{\rm SER} = 0.56(1)$ (blue square), where the results start to significantly deviate from the linear relation. (d) The power-law fit (line) of $\tau_{\rm D}$ in the SER regime gives  $\varphi_{\rm d} = 0.594(1)$ and $\gamma=1.6$.} 
\label{fig:phid3d}
\end{figure}

We follow the strategy developed in Ref.~\cite{Charbonneau2014} to determine the location of the dynamical (mode-coupling theory -- MCT) crossover $\varphi_{\rm d}$. (i) We obtain the diffusion time $\tau_D = \bar{\sigma}^2/D$, where $D$ is the long-time diffusivity and the average particle diameter, $\bar{\sigma}$, is also the unity of length. At long times, the mean-squared displacement (MSD) $\Delta (t) =
\frac{1}{N} \sum _{i=1}^{N} \av{|\vr_i(t ) - \vr_i(0)|^2}$ is dominated by the diffusive behavior $\Delta(t) = 2dDt = 2d \bar{\sigma}^2 (t/\tau_D)$ (Fig.~\ref{fig:phid3d}a). Note that we here ignore the dependence of $\Delta(t)$ on $t_{\rm w}$ (compared with Eq. (1) in the main text), because we are interested in equilibrium liquid states below $\varphi_{\rm d}$, where no aging is observed. (ii) We determine the structural relaxation time $\tau_\alpha$ by collapsing the mean-squared typical displacement (MSTD) $r^2_{\rm typ}(t/\tau_\alpha)$ in the caging regime (Fig.~\ref{fig:phid3d}b), where the typical displacement  $r_{\rm typ} (t)$ is defined as  $r_{\rm typ} (t) = \lim_{z \to 0} 
\frac{1}{N} \sum _{i=1}^{N} \av{|\vr_i(t ) - \vr_i(0)|^z}^{1/z}$. (iii) We find the  
density threshold $\varphi_{\rm SER} = 0.56(1)$ for the breakdown of Stokes-Einstein relation (SER), $D \propto \eta^{-1}$, where $\eta$ is the shear viscosity. Because $\tau_D \propto 1/D$ and $\tau_\alpha \propto \eta$ in this regime, the SER can be rewritten as $\tau_D \sim \tau_\alpha$ (Fig.~\ref{fig:phid3d}c). (iv) We fit the time $\tau_D$ in the SER regime ($\varphi < \varphi_{\rm SER}$) to the MCT scaling
$\tau_D \propto |\varphi - \varphi_{\rm d}|^{-\gamma}$ (or equivalently, $D \propto |\varphi - \varphi_{\rm d}|^{\gamma}$)
to extract $\varphi_{\rm d}= 0.594(1)$ (Fig.~\ref{fig:phid3d}d).
\newline

\section{Decompression of equilibrium configurations above the dynamical crossover $\varphi_{\rm d}$}
\begin{figure}[h]
\centerline{\hbox{ \includegraphics [width = 0.7\columnwidth] {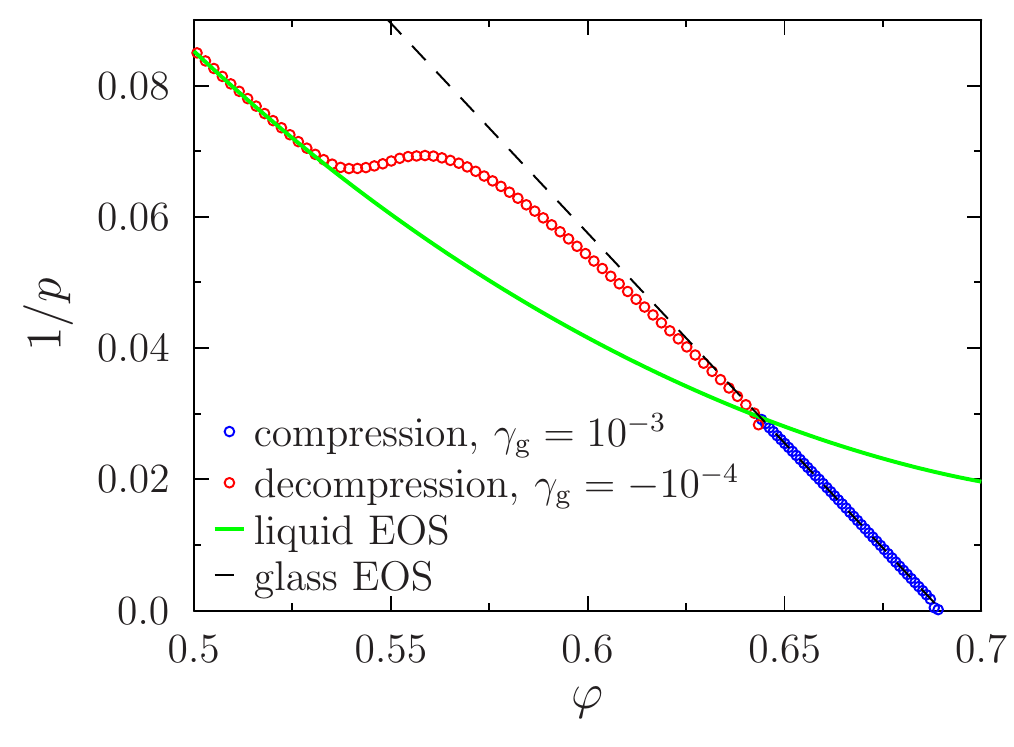}}}
\caption{Compression and decompression (negative $\gamma_{\rm g}$) of an initial equilibrium configuration at $\phi_{\rm g} = 0.643$.}
\label{fig:decomp3d}
\end{figure}

The equilibrium liquid configurations obtained from the Monte-Carlo swap algorithm are in the deeply supercooled regime $\varphi_{\rm g} > \varphi_{\rm d}$, where the structural $\alpha$-relaxation and thus diffusion are both strongly suppressed. As long as $\varphi_{\rm g}$ is sufficiently far beyond $\varphi_{\rm d}$, the MSD for $\varphi \geq\varphi_{\rm g}$ exhibits a well-defined plateau, and the diffusive regime is not observed in the MD simulation window (see Figs. 2a and 2b in the main text). To further reveal the separation between the $\alpha$- and $\beta$-relaxations, we decompress the equilibrium configuration and show that 
the resulting equation of state (EOS) follows the free-volume glass EOS ( Eq. (3) in the main text) up to a threshold density, at which the system melts into a liquid (Fig.~\ref{fig:decomp3d}). This behavior suggests that our compression/decompression is slower than the $\beta$-relaxation and much faster than the $\alpha$-relaxation, such that the system is kept  within a glass state. If the $\alpha$-relaxation were faster than the decompression, the state would follow the liquid EOS instead of the glass EOS under decompression. Note that a similar phenomenon has been reported in simulations of ultrastable glasses~\cite{Singh2013, Hocky2014}.

\begin{figure}[h]
\centerline{\hbox{ \includegraphics [width = \columnwidth] {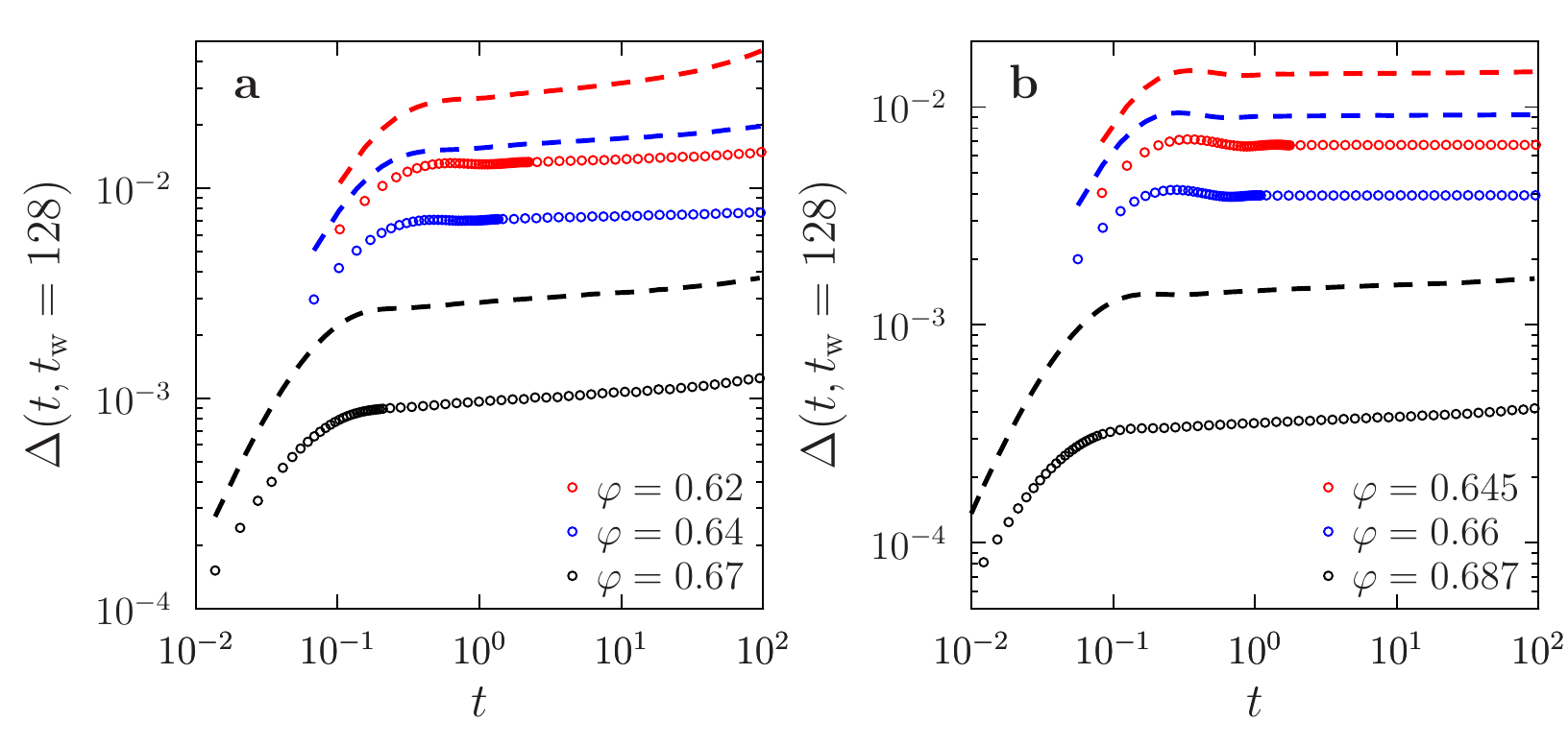}}}
\caption{The MSD $\Delta(t,t_{\rm w}=128)$ of larger (circles) and smaller (dashed lines) half particles, for (a) $\varphi_{\rm g} = 0.619$ and (b) $\varphi_{\rm g} = 0.643$.}
\label{fig:sizep}
\end{figure}
\begin{figure}[h]
\centerline{\hbox{ \includegraphics [width = 0.7\columnwidth] {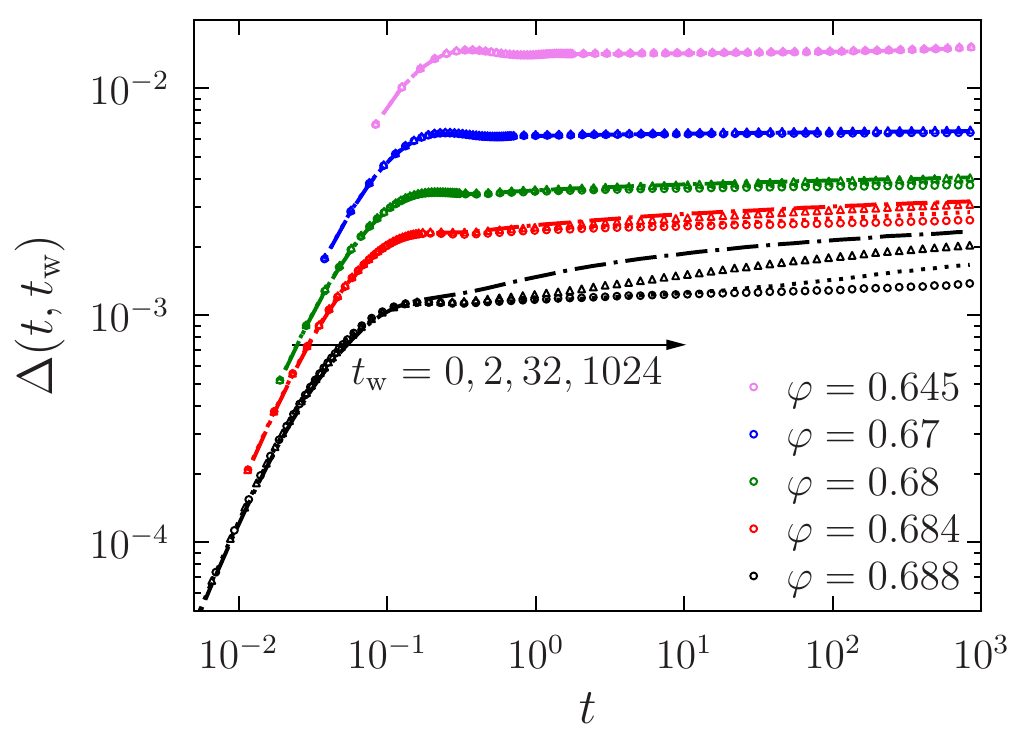}}}
\caption{Time evolution of $\Delta (t, t_{\rm w})$ of the smaller half of the particles  for several $t_{\rm w}$ and $\varphi$, following compression from $\varphi_{\rm g} = 0.643$.}
\label{fig:aging_smallp}
\end{figure}

\section{Particle size effects}
Suppressing the $\alpha$-relaxation and diffusion is crucial to our analysis. Besides pushing $\varphi_{\rm g}$ to higher densities, we find that it is useful to filter out the contribution of smaller particles, which are usually more mobile, from the calculation of the observables. For example,  the MSD of the smaller half of the particle size distribution grows faster and diffuses sooner than that of the larger half (Fig.~\ref{fig:sizep}). 
The diffusion of smaller particles,

however, vanishes as $\varphi_{\rm g}$ increases, which suggests that the effect is not essential to the underlying physics but an artifact of our choice of system.  For example, Fig.~\ref{fig:aging_smallp} shows that the smaller particles have very similar aging behavior as the larger particles (compared with Fig. 2a).
For this reason, $\Delta(t, t_{\rm w})$ and $\Delta_{AB}(t, t_{\rm w})$ in this work are always calculated using only the larger half of the particle distribution.

\section{Distribution of single particle cage sizes}
\begin{figure}[h]
\centerline{\hbox{ \includegraphics [width = 0.6\columnwidth] {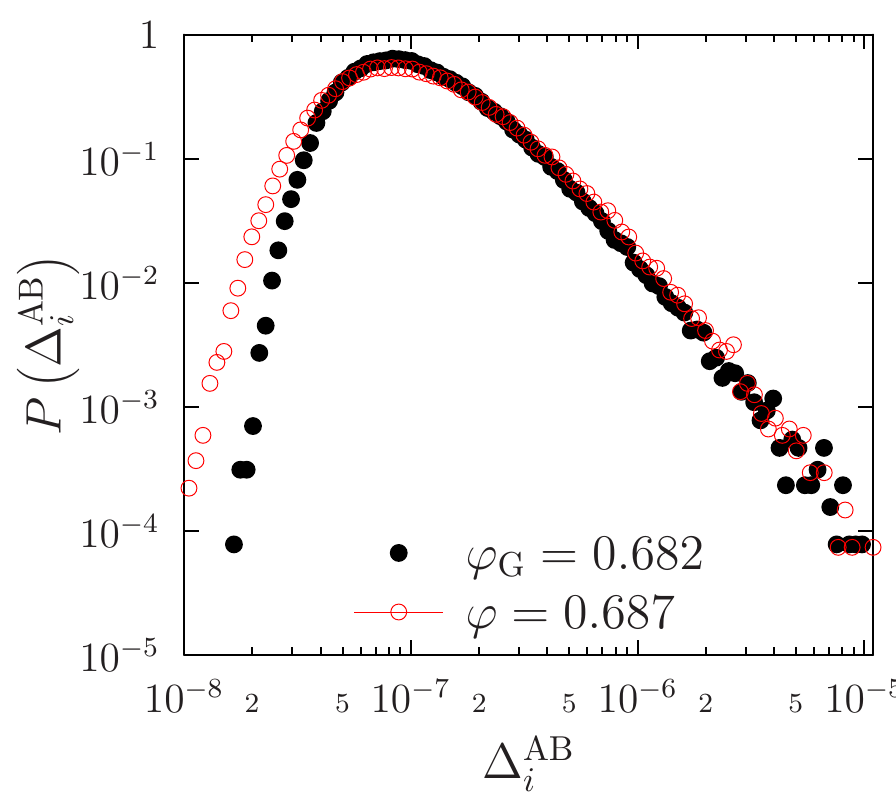}}}
\caption{  Distribution of particle cage sizes near the Gardner point, $\phi_\mathrm{G}=0.682$, and above, $\phi=0.687$, for $N=8000$ and $\varphi_\mathrm{g}=0.640$. }
\label{fig:ratlers}
\end{figure}
It is well known that, at the jamming point
  in finite dimensions, not all particles are part of
  the mechanically rigid network. 
  Particles that are excluded from this network rattle relatively freely within their empty
  pores, hence the name
  ``rattlers''. Because these localized excitations are not included in
  the infinite-dimensional theory, one could wonder whether the
  particles destined to become rattlers at jamming might play a role in
  our determination of the Gardner transition at finite
  dimensions. We argue here that it is not the case. Indeed,
  as shown in previous studies (see e.g.~\cite{CCPZ12}),
  the effect of rattlers becomes important only for reduced pressure $p \gtrsim10^4$. The Gardner line detected in this work covers much
  lower reduced pressures, $30 \lesssim p_{\rm G}
  \lesssim 500$, which allows us to ignore rattlers.

  Further support for this claim can be obtained by considering the
  probability distribution function of individual particle cage sizes
  $\Delta_i^{AB}=\left\langle|\V{r}_i^A-\V{r}_i^B|^2\right\rangle_\mathrm{th}$
  calculated from many samples 
  (with $\left\langle \ldots \right \rangle_\mathrm{th}$ the thermal averaging)
  at the
  Gardner point and above it (Fig.~\ref{fig:ratlers}), for one of the
  densest $\varphi_\mathrm{g}$. Both distributions show a single peak
  with a power-law tail (which is consistent with previous
  work~\cite{maiti2014free}). If rattlers gave a second peak, then they should be removed in our analysis, but this is not the case here.

\section{Caging timescale}

\begin{figure}[h]
\centerline{\hbox{ \includegraphics [width = \columnwidth] {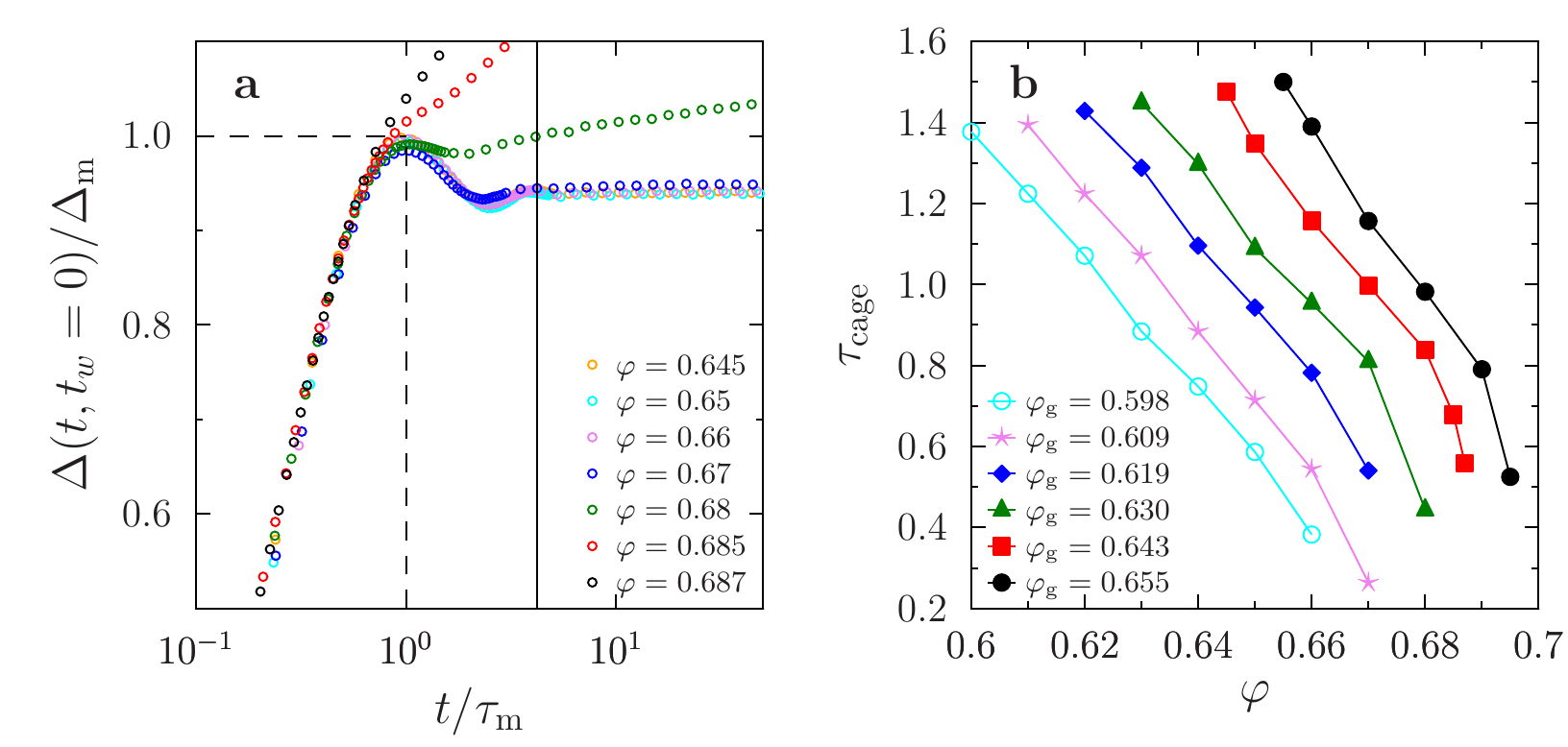} }}
\caption{(a) Rescaled MSD, $\Delta(t, t_{\rm w}=0)$, for $\varphi_{\rm g} = 0.643$, where the solid vertical line corresponds to the caging onset time $\tau_{\rm cage}$. (b) Dependence of $\tau_{\rm cage}$ on $\varphi$ for different $\varphi_{\rm g}$.}
\label{fig:tau0}
\end{figure}

We define a timescale $\tau_{\rm cage}$ to characterize the onset of caging.
The ballistic regime of MSD at different $\varphi$ is described by a master function $\Delta(t, t_{\rm w})/\Delta_{\rm m}\sim \Delta_{\rm ballistic} (t/\tau_{\rm m})$, independent of waiting time $t_{\rm w}$, where the microscopic parameters  $\tau_{\rm m}$ and $\Delta_{\rm m}$ correspond to the peak of the MSD (see Fig.~\ref{fig:tau0}a). To remove the oscillatory peak induced by the finite system size, we introduce $\tau_{\rm cage}$ 
 slightly larger than, but proportional to
$\tau_{\rm m}$, so that $\tau_{\rm cage}$ corresponds to the beginning of the plateau. The same collapse is obtained for any $t_{\rm w}$; as a result, $\tau_{\rm cage}$ is independent of $t_{\rm w}$.
Above $\varphi_{\rm G}$, $\tau_{\rm cage}$ is the time needed for relaxing the fastest vibrations. 
The dependence of $\tau_{\rm cage}$ on $\varphi_{\rm g}$ is summarized in Fig.~\ref{fig:tau0}b. Note that $\tau_{\rm cage} \sim {\cal O}(1)$
with weak variation for all considered state points.

By contrast, the mean-squared distance between two copies, $\Delta_{AB}(t)$, depends only weakly on $t$ (see Fig. 2b in the main text). Our choice of $t=\tau_{\rm cage}$ therefore does not affect substantially the value of  $\Delta_{AB}  \equiv  \Delta_{AB}(\tau_{\rm cage})$ (Fig. 3 in the main text). Note that, $\Delta_{AB}$ basically describes the asymptotic long time behavior of $\Delta(t)$, i.e,  $\Delta_{AB}  \approx \Delta_{AB}(t \to \infty) \simeq \Delta(t \to \infty, t_{\rm w} \to \infty)$.

\section{Absence of crystallization and of thermodynamic anomalies at the Gardner density}

 It is quite obvious that, because the system is not diffusing away
 from the original liquid configuration at $\f_{\rm g}$ during the
 simulation time window, no crystallization can happen in the glass
 regime.  Indeed, when crossing $\varphi_{\rm G}$ no sign of incipient
 crystallization or formation of comparable anomaly appears in the
 pair correlation function.  Also, $\frac{d (1/p)}{d\f}$ is
 essentially constant in the glass regime; nothing special happens to
 this quantity at $\varphi_{\rm G}$. The crossover would thus remain
 invisible if we only considered the compressibility and not more
 sophisticated observables.

 \section{Time evolution of
     $\delta\Delta(t,t_\mathrm{w})$ and timescales}
\begin{figure}[h]
\centerline{\hbox{ \includegraphics [width = 1.0\columnwidth] {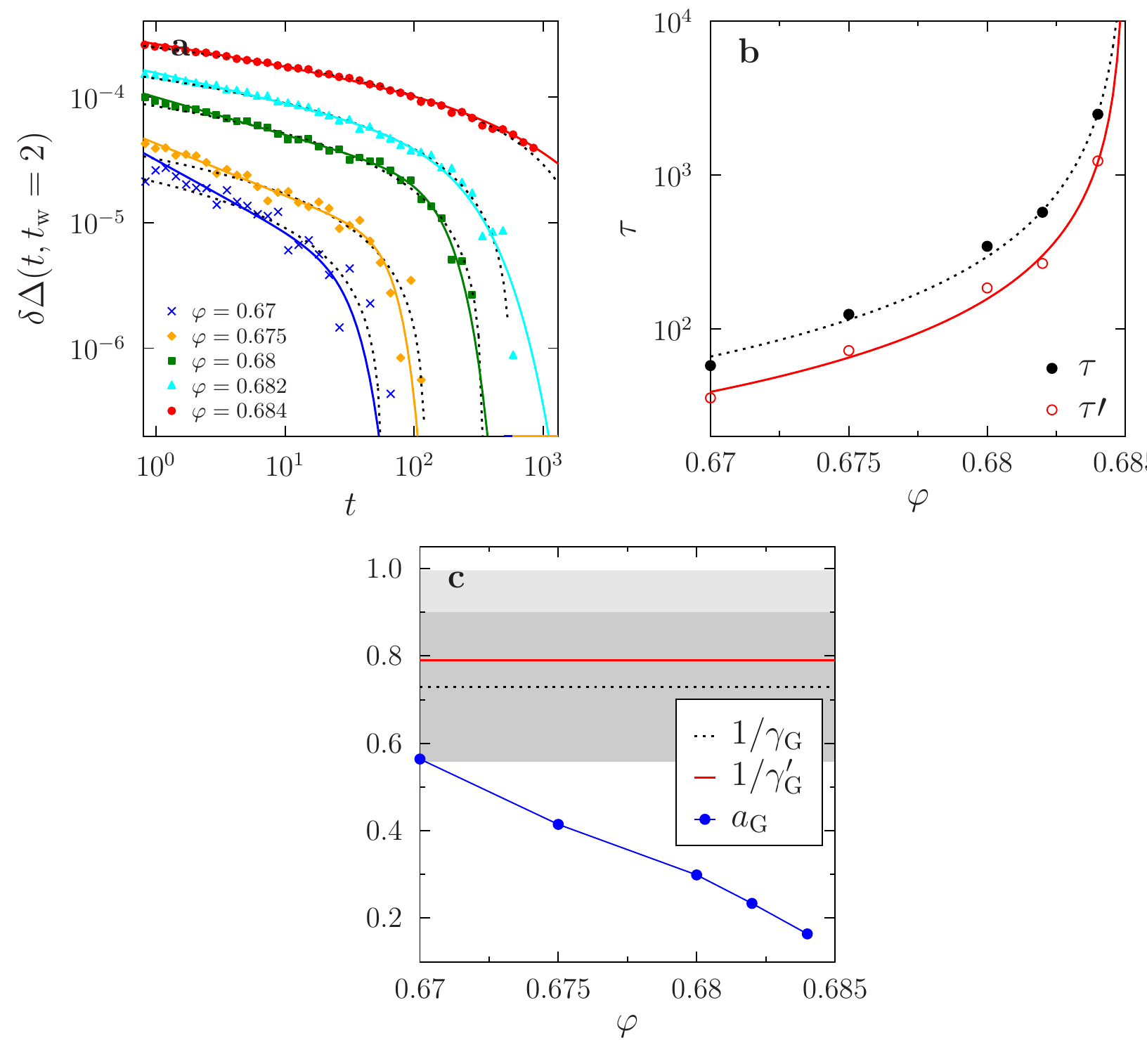}}} 
\caption{ (a) Time evolution of
    $\delta\Delta(t,t_\mathrm{w})$ for $t_\mathrm{w}=2$ and
    $\varphi_{\rm g}=0.643$. The data was fitted to
    $\delta\Delta(t,t_\mathrm{w})\sim t^{-a_{\rm G}}\exp[-(t/\tau^\prime)^{b_{\rm G}}]$
    (solid lines) and to the fitting form discussed in the main text,
    $\delta\Delta(t,t_\mathrm{w})\propto 1-\ln t/\ln \tau$ (dashed
    lines). The scalings provide a characteristic
    relaxation time, $\tau^\prime$ and $\tau$, respectively, for each $\varphi$. (b)
    The two estimates behave similarly
    around $\varphi_\mathrm{G}=0.684(1)$, and both can be fit
    to a
    power-law, $\tau\propto
    \left(\varphi_\mathrm{G}^\tau-\varphi\right)^{-\gamma_{\rm G}}$. (c)
    Comparison between $\gamma_{\rm G}^{-1}$ (gray
    gives the uncertainty region) and $a_{\rm G}$ obtained 
    from the scaling function Eq.~\eqref{eq:fitMCT}. Results appear 
    incompatible with the MCT prediction $a_{\rm G}=\gamma_{\rm G}^{-1}$,
    but a similar behavior was observed in
    a mean-field model~\cite{CJPRSZ2015PRE}. It may thus be due to the difficulty of fitting $a_{\rm G}$ in the critical regime.
    }
\label{fig:fitMCT}
\end{figure}

     The relaxation
   timescale $\tau$ (Fig. 2d in the main text) was extracted from
   fitting the long-time behavior of $\delta\Delta(t,t_\mathrm{w})$
   to a logarithmic scaling formula
   $\delta\Delta(t,t_\mathrm{w})\propto 1-\ln t/\ln \tau$, following
   the strategy discussed in previous
   works~\cite{CJPRSZ2015PRE,janus:campodyn}. This particular choice of 
   functional form makes it easier
   to obtain reliable fits for the whole window of parameters
   $t_\mathrm{w}$ and $\varphi$. A more
   conventional scaling function, such as
\begin{equation}\label{eq:fitMCT}\delta\Delta(t,t_\mathrm{w})\sim
  t^{-a_{\rm G}}\exp\left[-\left(\frac{t}{\tau^\prime}\right)^{b_{\rm G}}\right],\end{equation}
has a stronger theoretical motivation, but at the cost of requiring more fitting constants.
Data is nonetheless also well described by this functional form, as we
show in Fig.~\ref{fig:fitMCT}a, where the solid lines are obtained
from fits to the functional form Eq.~\eqref{eq:fitMCT} and the dashed
lines are the logarithmic fits from Fig. 2c in the main
text.

One can also fit Eq.~\eqref{eq:fitMCT} to extract a new estimate of
the timescale $\tau^\prime$. 
If doing so, we obtain results essentially proportional
to our previous estimate of $\tau$ (Fig.~\ref{fig:fitMCT}b).
In this sense, both scalings appear to be equivalent (at least for low $\varphi_\mathrm{g}$ and $t_\mathrm{w}$ where both fits can be done). More
quantitatively, one can also attempt a fit of the two characteristic
times to a power-law divergence,
\begin{equation}\label{eq:tau}
\tau\propto \left(\varphi_\mathrm{G}^\tau-\varphi\right)^{-\gamma_{\rm G}},
\end{equation}
 as is expected in the vicinity of a critical point (Fig.~\ref{fig:fitMCT}b). 
 Again, we obtain values within
 the error bars of the parameters. In particular,
 $\gamma_{\rm G}=1.3(3)$ and $\varphi_\mathrm{G}^\tau=0.685(1)$, 
 which is completely compatible with our best estimate for the Gardner
 point, $\varphi_{\rm G} = 0.684(1)$, at this $\varphi_{\mathrm{g}}$.

MCT further suggests that there should be a relation $\gamma_{\rm G}=1/a_{\rm G}$ between
the exponent $\gamma_{\rm G}$ obtained from the fit of Eq.~\eqref{eq:tau} and
the exponent $a_{\rm G}$ from the fit of Eq.~\eqref{eq:fitMCT}. At this point,
we could not fit the data using a constant value of $a_{\rm G}$ for all 
densities. Instead, the exponent decreases 
as $\varphi$ increases (see Fig.~\ref{fig:fitMCT}c), and becomes 
increasingly incompatible with $\gamma_{\rm G}$ extracted from the
fit Eq.~\eqref{eq:tau}. Although this fact seems to be inconsistent with the
mean-field theory, the same behavior of $a_{\rm G}(\varphi)$ (also quantitatively) was recently reported in a similar study in a
mean-field model of hard spheres (HS) over comparable timescales~\cite{CJPRSZ2015PRE}. This
inconsistency might thus be related to the inner technical difficulty of
fitting the exponent $a$ using Eq.~\eqref{eq:fitMCT}.

\section{Time dependence of the skewness and determination of the Gardner density}
In this section, we explicitly show that the position of the peak of the skewness $\Gamma_{AB}$, which is used to extract the location of the Gardner point $\varphi_{\rm G}$  (Fig. 3d in the main text), is independent of the choice of timescale, $\tau_\mathrm{cage}$. Promoting the skewness to a time-dependent quantity, $\Gamma_{AB}(t)=\av{w_{AB}^3(t)}/\av{w_{AB}^2(t)}^{3/2}$, with $w_{AB}(t)=\Delta_{AB}(t)-{\av{\Delta_{AB}(t)}}$,  
confirms that the peak position of $\Gamma_{AB}(t)$ is nearly invariant of $t$, although the peak height does have a small time dependence. 

\begin{figure}[h]
\centerline{\hbox{ \includegraphics [width = 0.7\columnwidth] {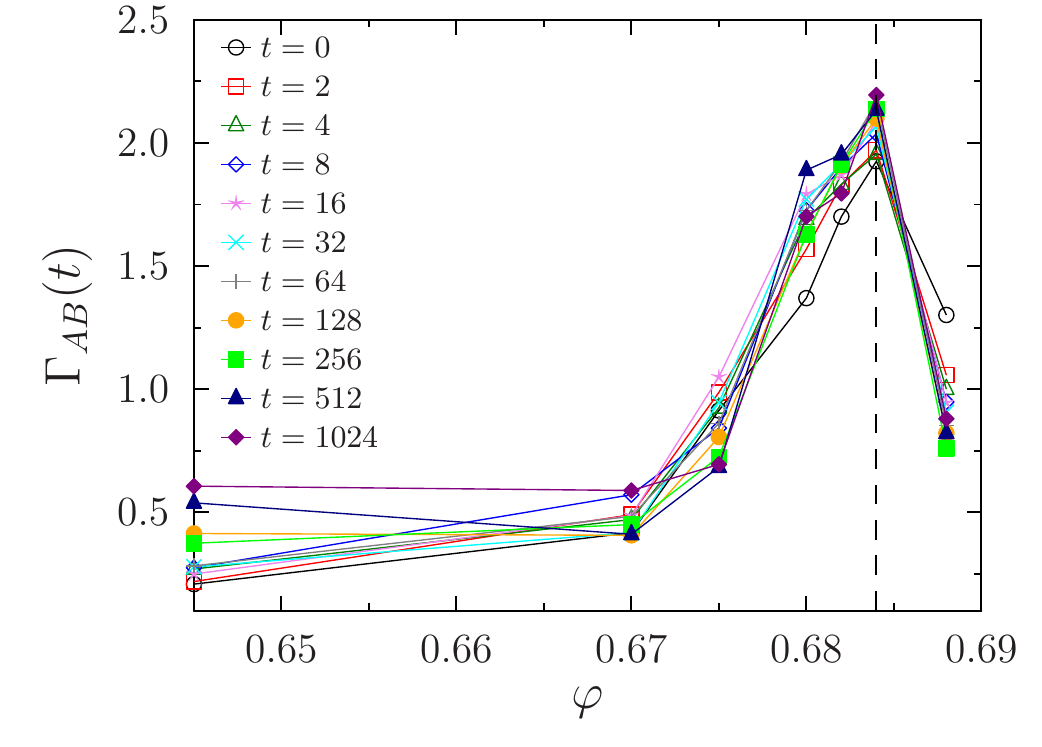}}} 
\caption{Time dependence of the caging skewness $\Gamma_{AB}(t)$ for $\varphi_{\rm g} = 0.643$. The peak position, which gives $\varphi_{\rm G}$ (dashed vertical line), is invariant of $t$.}
\label{fig:skew_time}
\end{figure}

\section{System-size dependence of the susceptibility at the Gardner transition}
\begin{figure}[h]
\centerline{\hbox{ \includegraphics [width = 0.7\columnwidth] {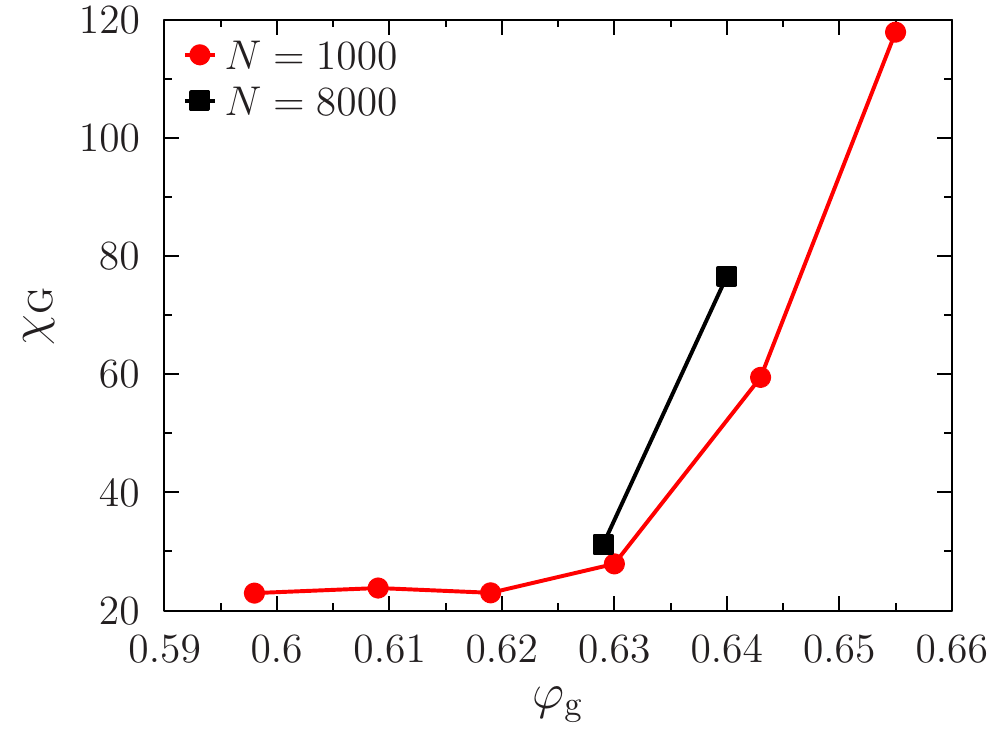}}} 
\caption{ System size dependence of the caging susceptibility $\chi_{\rm G}$ at the Gardner transition, where $\chi_{\rm G}$  is plotted as a function of the initial density $\varphi_{\rm g}$ for two different system sizes.}
\label{fig:chiG}
\end{figure}

From a theoretical viewpoint, whether the mean-field Gardner
transition persists in finite dimensions is still under
debate~\cite{UB14}. In this work, we have shown the existence of a
crossover (reminiscent of the mean-field Gardner transition) at two
system sizes, $N=1000$ and $N=8000$. However, the proof of the
existence of the Gardner transition in the thermodynamic limit would
require a systematic use of finite-size scaling
techniques~\cite{amit2005field}, which is beyond the scope of this
 paper. Previous studies further suggest that this kind of
analysis might be extremely challenging. For example, symmetry arguments
suggest that the Gardner transition should be in the same universality
class as the de Almeida-Thouless line in mean-field spin-glasses in a
field~\cite{UB14}, whose finite-dimensional persistence is still the object of
active debate even after intensive numerical
scrutiny~\cite{young:04,larson:13,baity:14,banos:12}.  
One way to test whether our data are compatible with a true
phase transition is by checking that the
caging susceptibility at the transition point, $\chi_{\rm G} \equiv
\chi_{AB}(\varphi = \varphi_{\rm G})$, appears to divergence at  $N\to\infty$. Considering that $\chi_{\rm G}$ must be finite in a
finite system (as shown it Fig. 3c for $N=1000$), the divergence
requires that $\chi_{\rm G}$ increases with $N$. We can see that this requirement is fulfilled in Fig.~\ref{fig:chiG}, which compares susceptibilities for $N=1000$ and 8000.

\section{Compression-Rate Dependence}
\begin{figure}[h]
\centerline{\hbox{\includegraphics [width = 1\columnwidth] {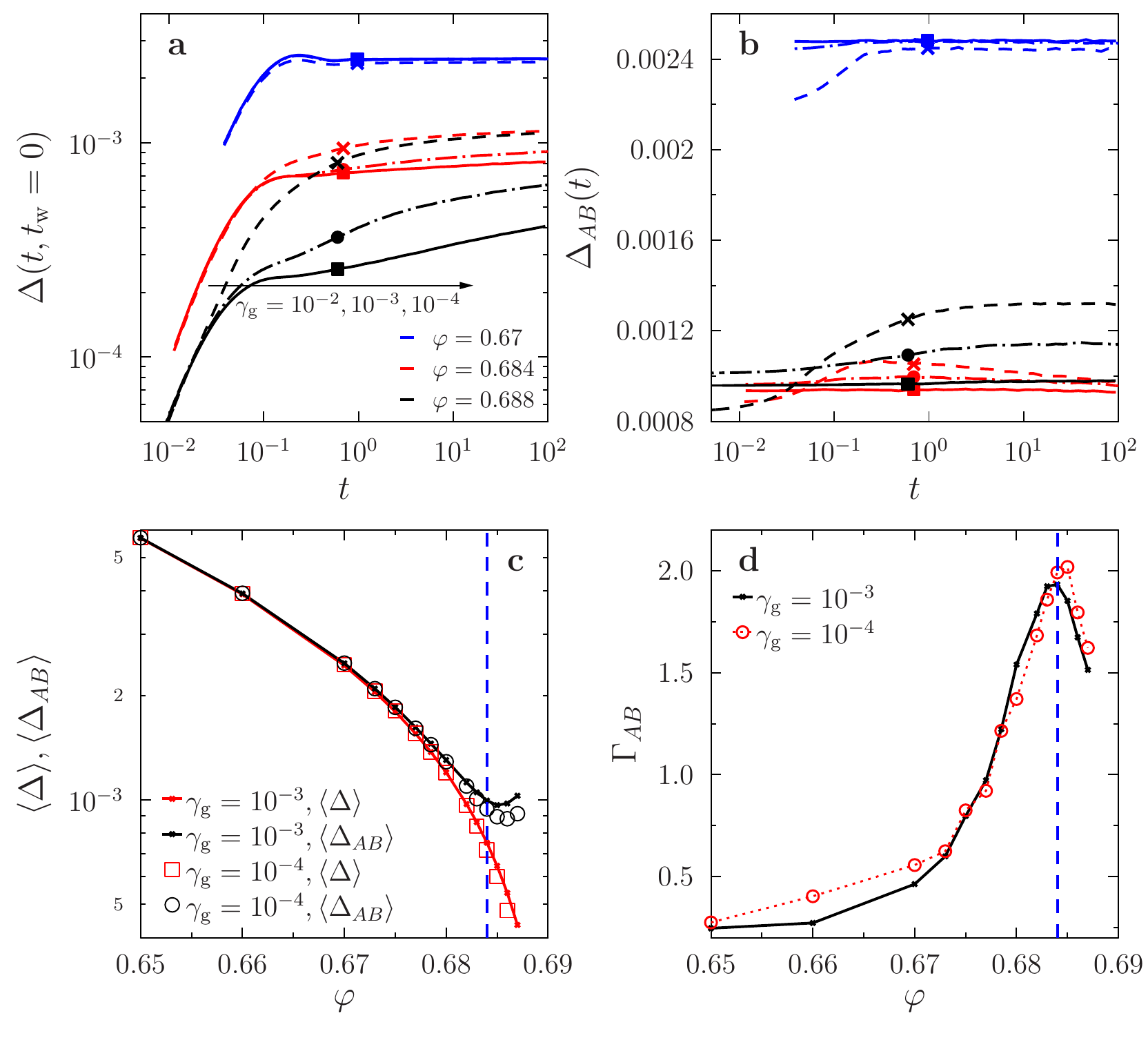}}}
\caption{ Compression-rate dependence. (a-b) MSD $\Delta(t, t_{\rm w} = 0)$ and $\Delta_{AB}(t)$ for different compression rates and densities for $\varphi_{\rm g} = 0.643$.  The mean caging order parameters $\langle \Delta \rangle$ and $\langle \Delta_{AB} \rangle$ are marked by crosses, circles, and squares for $\gamma_{\rm g}=10^{-2}$, $10^{-3}$ and $10^{-4}$, respectively. (c) For $10^{-3} \leq \gamma_{\rm g} \leq 10^{-4}$, $\langle \Delta \rangle$ and $\langle \Delta_{AB} \rangle$ are nearly independent of $\gamma_{\rm g}$ when $\varphi < \varphi_{\rm G}$ ($\varphi_{\rm G}$ is denoted by the dashed blue line), and weakly dependent on  $\gamma_{\rm g}$ when $\varphi > \varphi_{\rm G}$.  (d) The peak position of the skewness, $\varphi_{\rm G} = 0.684(1)$, is independent of $\gamma_{\rm g}$ within the numerical accuracy.}
\label{fig:rate_dependence}
\end{figure}

In this section, we consider how our overall analysis depends on the compression rate $\gamma_{\rm g}$ used for preparing samples. In principle, a proper $\gamma_{\rm g}$ should be such that particles have sufficient time to equilibrate their vibrations but not to diffuse. In other words, the timescale associated with compression, $\tau_{\rm g} \sim 1/\gamma_{\rm g}$, should lie between the $\alpha-$ and $\beta-$relaxation times,  $\tau_\beta < \tau_{\rm g} < \tau_\alpha$. For our system, we observe that when $10^{-3} \leq \gamma_{\rm g} \leq 10^{-4}$ and  $\varphi < \varphi_{\rm G}$, both $\Delta(t, t_{\rm w})$ and $\Delta_{AB}(t)$ reach flat plateaus that are essentially independent of $\gamma_{\rm g}$  (see Figs. ~\ref{fig:rate_dependence}a and b). Thus in this range of compression rates, restricted equilibrium within a glass state is reached, while keeping the $\alpha-$relaxation sufficiently suppressed.  When $\varphi > \varphi_{\rm G}$,  however, $\Delta(t, t_{\rm w})$ and $\Delta_{AB}(t)$ display $\gamma_{\rm g}$-dependent aging effects consistent with a growing timescale in the Gardner phase.
As a result, the order parameters $\Delta$  and $\Delta_{AB}$, which are defined at the time scale of $\tau_{\rm cage}  \sim {\cal O}(1)$,  slightly depend on $\gamma_{\rm g}$ when $\varphi \gtrsim \varphi_{\rm G}$  (see Fig. ~\ref{fig:rate_dependence}c). This  mild $\gamma_{\rm g}$-dependence has nonetheless relatively little impact on our analysis of $\varphi_{\rm G}$. In particular, the location of the peak position of the caging skewness, based on which we determine the value of $\varphi_{\rm G}$, is independent of $\gamma_{\rm g}$ within the numerical accuracy (see Fig. ~\ref{fig:rate_dependence}d). Note that $\gamma_{\rm g}^{-1}$ plays a role akin to the waiting time $t_{\rm w}$. 
Varying $\gamma_{\rm g}$ is thus equivalent to varying $t_{\rm w}$ (see Fig.~2a and Fig.~\ref{fig:rate_dependence}a).

\section{Spatial correlation functions and lengths}
In this section, we provide a more detailed presentation of the spatial correlations of individual cages and of their associated length scales.  We consider both point-to-point and line-to-line spatial correlation functions, and show that the characteristic lengths obtained from both grow consistently around $\varphi_\mathrm{G}$. We also highlight the advantage of the line-to-line correlation function, which is used in the main text. Note that the results presented here are all for $N=1000$, while
in the main text results for the line-to-line correlation are reported for $N=8000$.

\begin{figure}[h]
\centerline{\hbox{\includegraphics [width = \columnwidth] {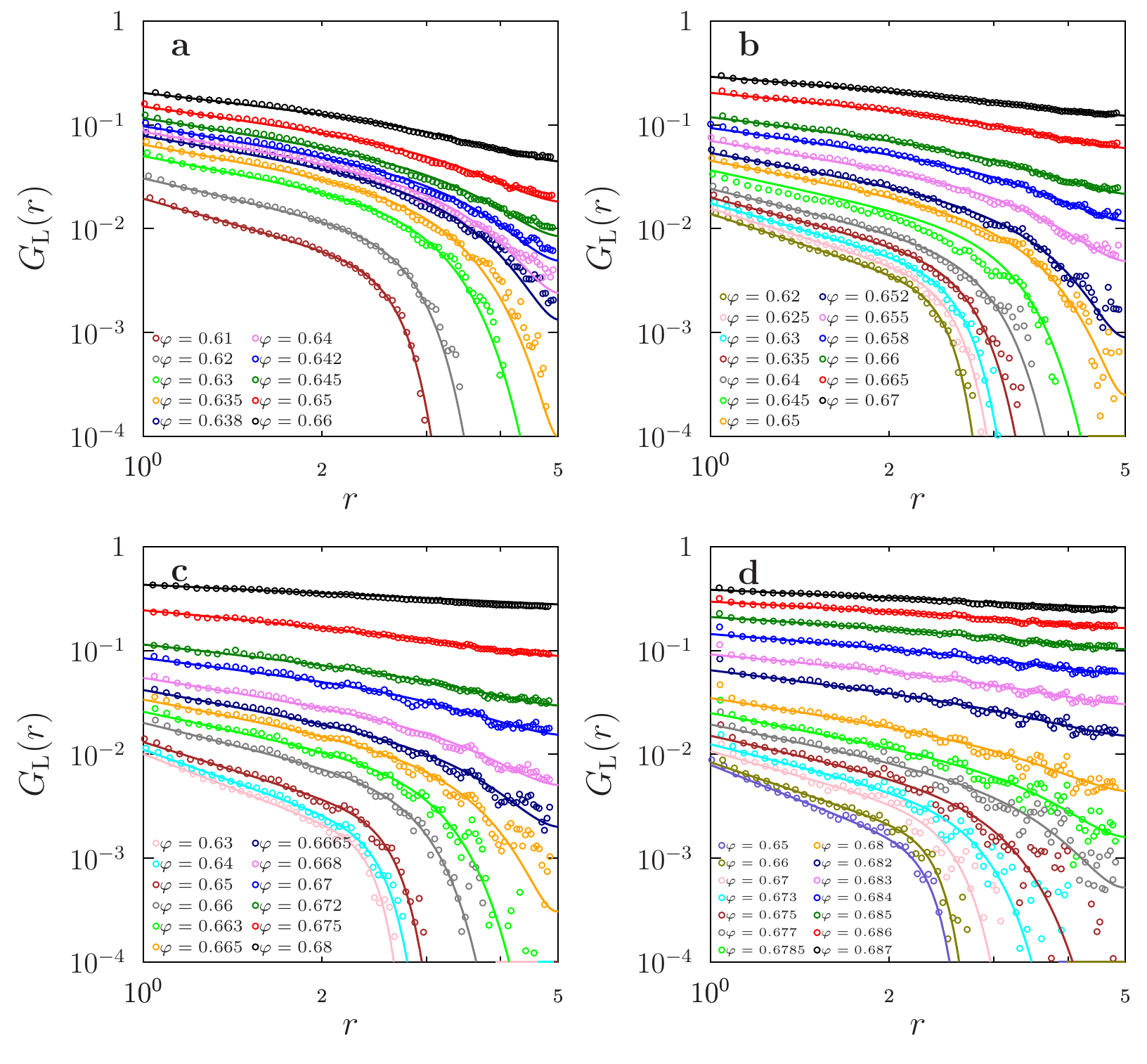}}}
\caption{The normalized line-to-line correlation functions $G_{\rm L}(r)$ are fitted to Eq.~(\ref{eq:GL}), for $\varphi_{\rm g}=0.609, 0.619, 0.630, 0.643$ (a-d).}
\label{fig:G}
\end{figure}

\begin{figure}[h]
\centerline{\hbox{\includegraphics [width = \columnwidth] {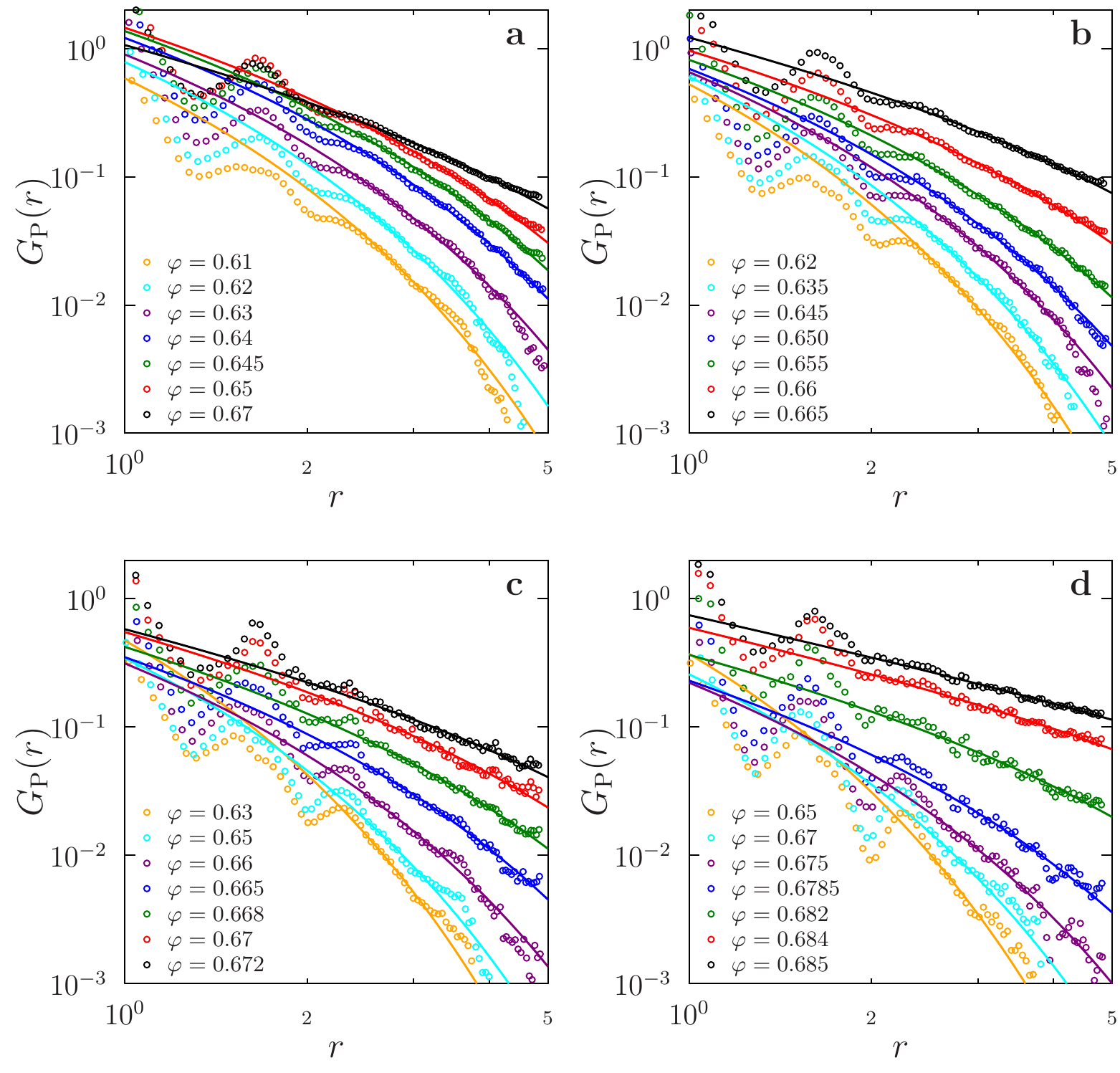}}}
\caption{The normalized point-to-point correlation functions $G_{\rm P}(r)$ are fitted to the form Eq.~(\ref{eq:GP}), for $\varphi_{\rm g}=0.609, 0.619, 0.630, 0.643$ (a-d).}
\label{fig:Gpoint}
\end{figure}

The susceptibility $\chi_{AB} = N \frac{ \av{\Delta^2_{AB}}-\av{\Delta_{AB}}^2}{\av{\Delta_{AB}}^2}$ discussed in the main text is directly associated with the unnormalized point-to-point spatial correlation function computed between two 
copies, $A$ and $B$, 
\begin{eqnarray}
G_{\rm P}^0(\V{r})=\frac{1}{N}\av{\sum_{i\neq j} u_i u_j\delta\left(\V{r}-|\V{r}_i^A-\V{r}_j^B|\right)},\nonumber\\\ \text{with}\ u_i=\frac{ |\V{r}_i^A-\V{r}_i^B|^2}{\av{\Delta_{AB}}}-1.
\end{eqnarray}
This definition gives $\int\mathrm{d}\V{r}\ G_{\rm P}^0(\V{r})=\chi_{AB}$.  Because in an 
isotropic fluid $G_{\rm P}^0(\V{r})$ is a rotationally invariant function, we define the normalized radial correlation
\beq
G_\mathrm{P}(r)=\frac{\av{\sum_{i \neq j} u_i
    u_j\delta\left(r-|\V{r}_i^A-\V{r}_j^B|\right)}}{\av{\sum_{i \neq j}\delta\left(r-|\V{r}_i^A-\V{r}_j^B|\right)}},
\label{eq:normalization}
\eeq
where  $r=|\V{r}|$ and the denominator is essentially the pair-correlation function between two clones,
\begin{equation}
g(r)=\frac{V}{N(N-1)}\av{\sum_{i \neq j} \delta\left(r-|\V{r}_i^A-\V{r}^B_j|\right)}.
\end{equation}
In a similar way, we define the normalized line-to-line spatial correlation function 
\beq G_{\rm L}(r) = \frac{\av{\sum_{\mu =
      1}^{3} \sum_{i\neq j}u_i
    u_j\delta\left(r-|\V{r}_{i,\mu}^A-\V{r}_{j,\mu}^A|\right)}}{\av{\sum_{\mu
      = 1}^{3} \sum_{i\neq j}
    \delta\left(r-|\V{r}_{i,\mu}^A-\V{r}_{j,\mu}^A|\right)}},
\label{eq:GL2}
\eeq
where $\V{r}_{i,\mu}$ is the projection of 
the particle position along the direction $\mu$. This last definition is also  Eq. (6) in the main text.

Both spatial correlation functions should capture the growth of vibrational heterogeneity around $\varphi_\mathrm{G}$ and are expected to decay at long distances as
\begin{equation}\label{eq:GlongrSI}
G(r)\to \frac{1}{r^a}F\left(\frac{r}{\xi}\right),
\end{equation}
where the damping function, $F(x)$, could in principle be different at large $x$ for $G_{\rm P}(r)$ and $G_{\rm L}(r)$. The function $F(r/\xi)$ is normally assumed to have an exponential or a stretched exponential form~\cite{Belletti2008PRL}. 
Equation~\eqref{eq:GlongrSI} suggests that $\chi_{AB}=4\pi\int \mathrm{d}r\ r^{2-a} F(r/\xi)\propto \xi^{3-a}$, which means that the observed growth in $\chi_{AB}$ around $\varphi_\mathrm{G}$ (Fig. 3c in the main text) should also be observed for 
$\xi$, but with a different exponent.

Although $G_\mathrm{P}(r)$ is the commonly used spatial correlation function, we find that $G_\mathrm{L}(r)$ traditionally used in lattice field theories~\cite{rothe2005lattice} is
more convenient to extract $\xi$. To illustrate why, we plot $G_\mathrm{L}(r)$ in Fig.~\ref{fig:G} and $G_\mathrm{P}(r)$ in Fig.~\ref{fig:Gpoint}  for different $\varphi_{\rm g}$. Compared to $G_{\rm P}(r)$, the line-to-line correlation function $G_{\rm L}(r)$ has the following advantages: (i) oscillations at small $r$ are removed, which better reveals the power-law scaling $r^{-a}$; (ii) it is easier to incorporate the periodic boundary condition by simply adding a symmetric term $F\left(\frac{L-r}{\xi}\right)$ to \eqref{eq:Glongr}, where $L$ is the linear size of the system; 
and (iii) the tail of $G_\mathrm{L}(r)$ decays faster than that of $G_\mathrm{P}(r)$, which implies a better separation between the power-law regime,  $r^{-a}$, and the tail, $F\left(r/\xi\right)$. The tail of $G_\mathrm{L}(r)$ is indeed well described by a 
stretched exponential, while $G_\mathrm{P}(r)$ has a slower exponential decay.

\pagebreak
\begin{figure}[h]
\centerline{\hbox{\includegraphics [width =\columnwidth] {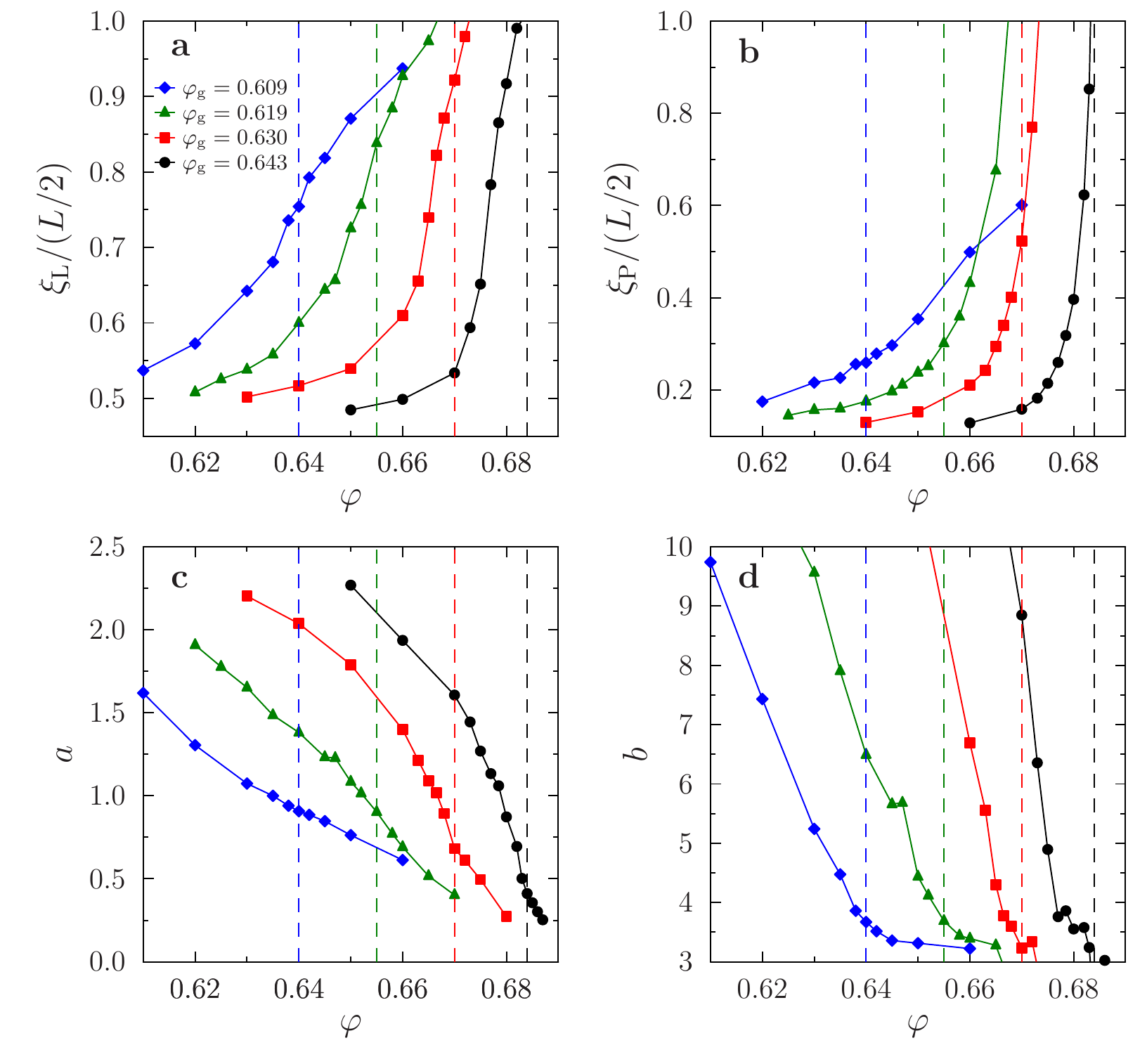}}}
\caption{(a and b) The correlation lengths $\xi_{\rm L}$ and $\xi_{\rm P}$ 
(in the main text $\xi=\xi_{\rm L}$)
as functions of $\varphi$ for a few different $\varphi_{\rm g}$. (c and d) The exponents (c) $a$ and (d) $b$ obtained from fitting $G_{\rm L}(r)$ with Eq.~\eqref{eq:GL}.}
\label{fig:xi}
\end{figure}

\begin{figure}[h]
 \centerline{\hbox{ \bf{(a)} \includegraphics[trim=300 10 400 50, clip,width =0.4\columnwidth]{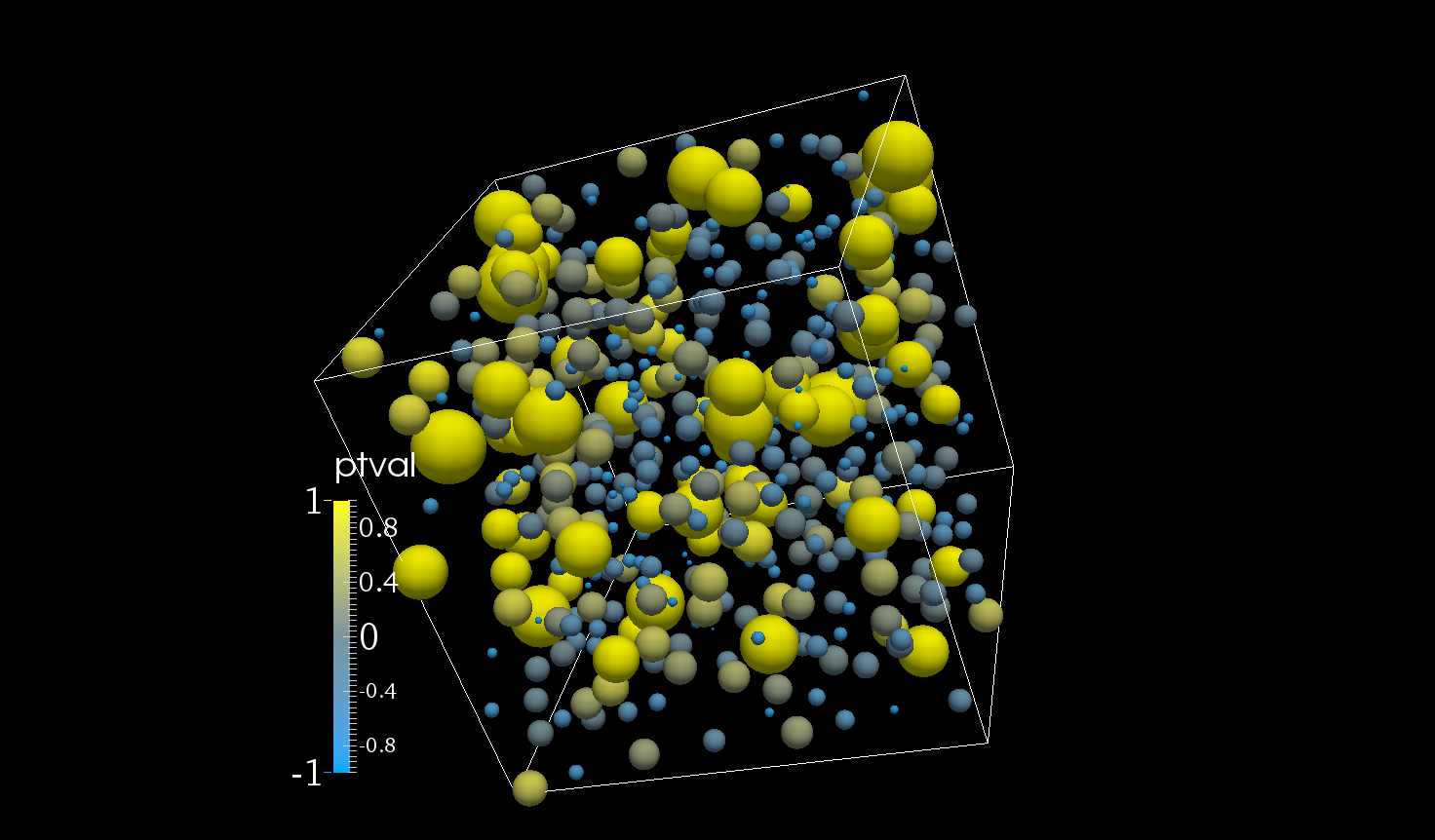} \bf{(b)}  \includegraphics[trim=300 10 400 50, clip,width = 0.4\columnwidth]{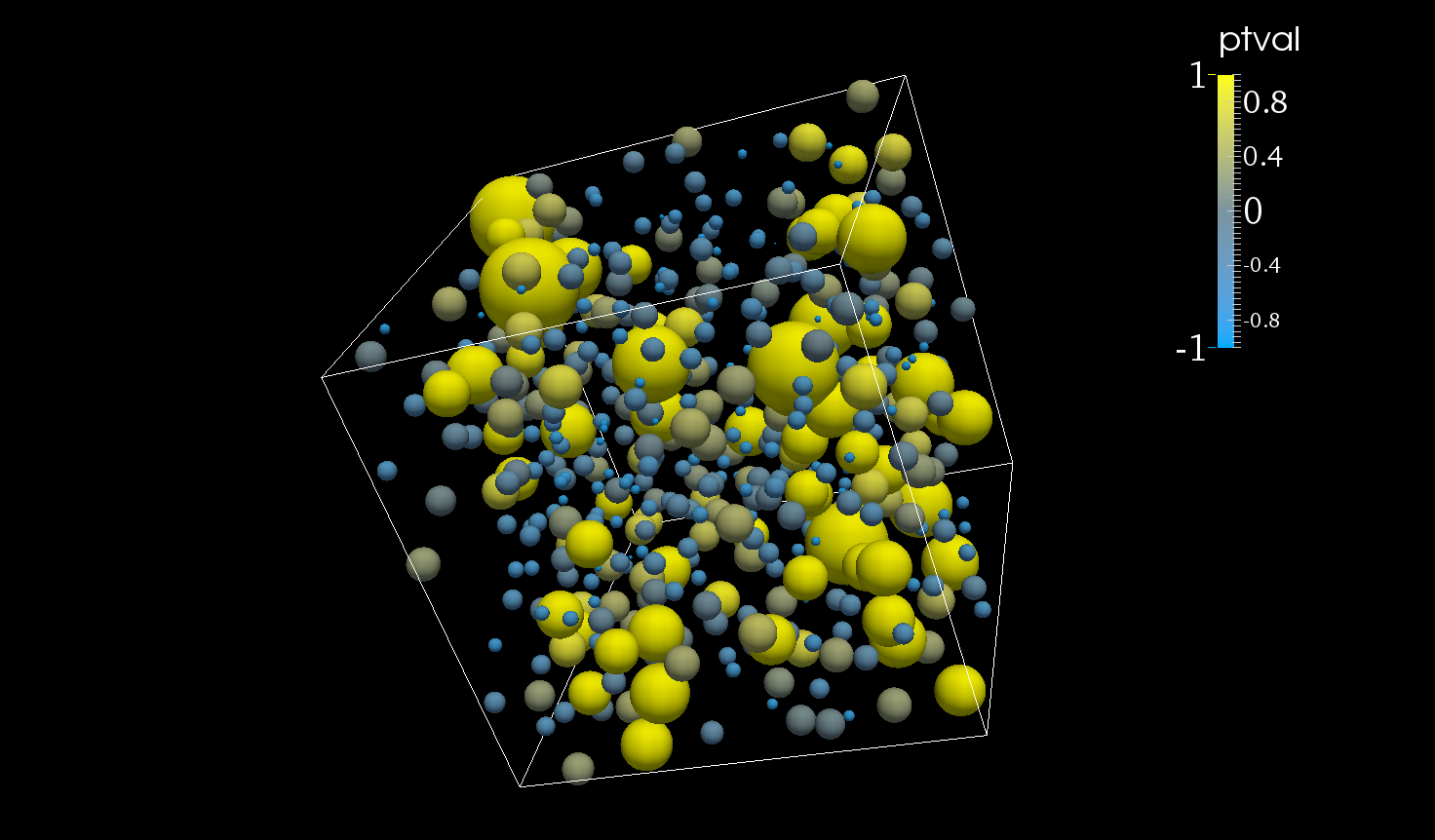}}}
  \centerline{\hbox{ \bf{(c)} \includegraphics[trim=300 10 400 50, clip,width = 0.4\columnwidth]{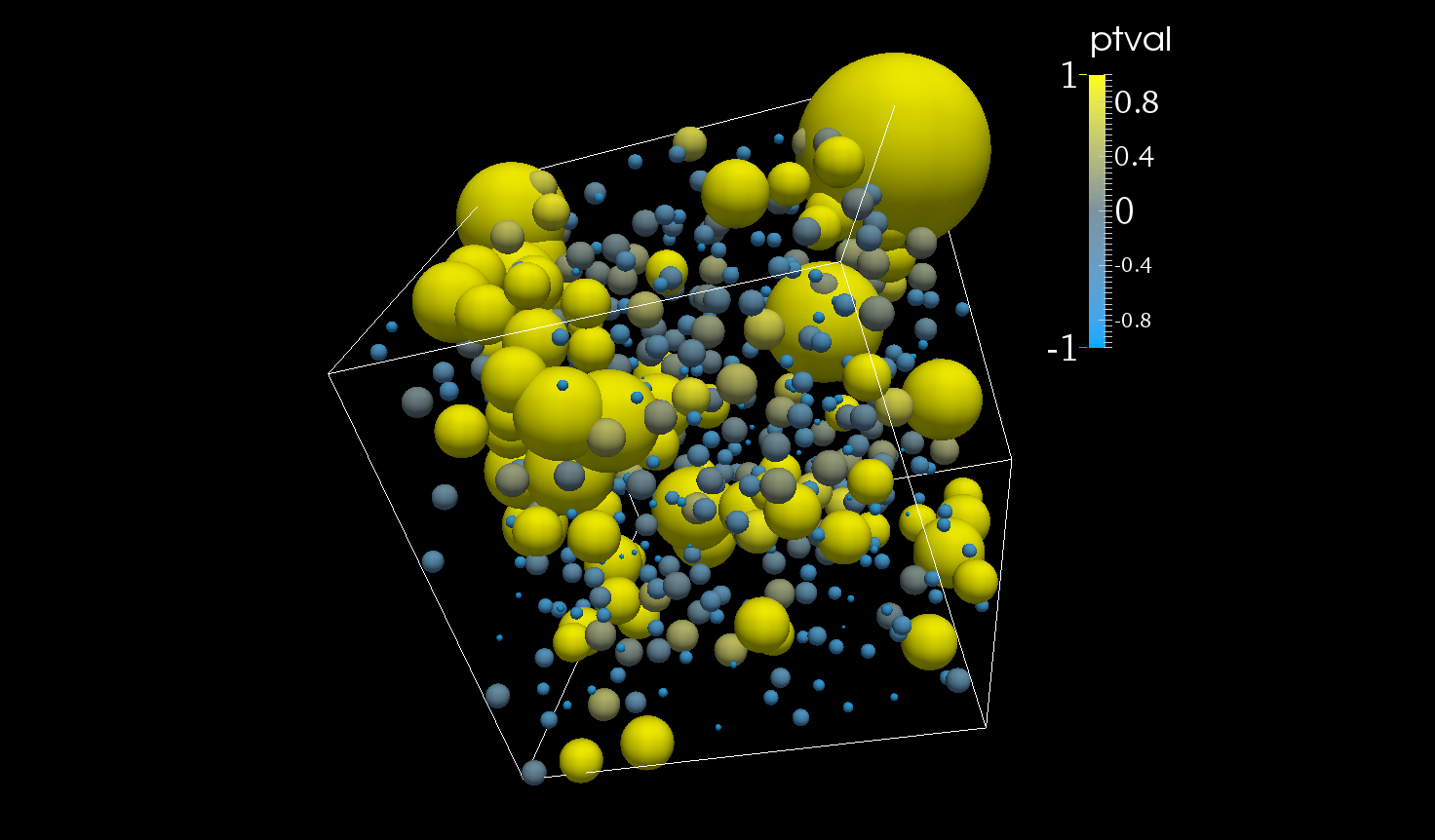} \bf{(d)}  \includegraphics[trim=300 10 400 50, clip,width = 0.4\columnwidth]{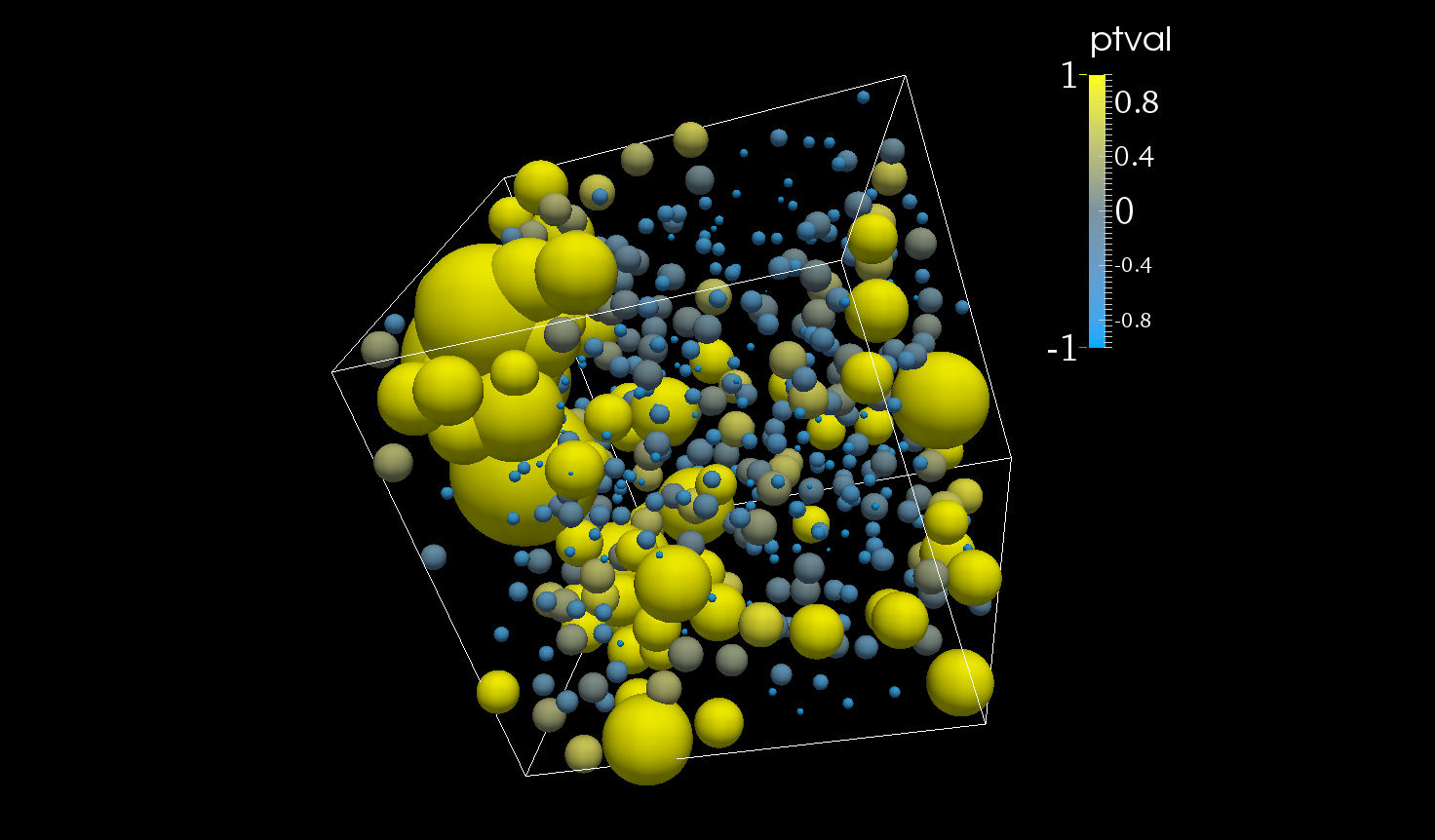}}}
 \caption{Snapshots of vibrational heterogeneity at different
  $\varphi$ (a-d, $\varphi = 0.63, 0.65, 0.67, 0.68$) for $\varphi_g=0.630$. Upon approaching
  $\varphi_\mathrm{G}=0.670(2)$, vibrations become increasingly
  heterogeneous.  The particle cages are represented as spheres
  centered at the $N$ particle positions of one of the simulated
  configurations. The sphere diameter at the
  position of the $i$-th particle, $\vr_i^A$, is proportional to
  $|\vr^A_i-\vr^B_i|$, the distance between its positions in two
  clones. The color stands for the deviation around the average
  $u_i=\frac{|\vr^A_i-\vr^B_i|^2}{\av{\Delta_{AB}}}-1$. 
  For the sake of visualization, all $u_i>1$ are plotted as $u_i=1$.}
\label{fig:snapshots}
\end{figure}

We extract $\xi_{\rm L}$ from fitting $G_\mathrm{L}(r)$ at different $\varphi$  (see Fig.~\ref{fig:xi}a) to the functional form  \begin{equation}\label{eq:GL}
G_\mathrm{L}(r)\sim r^{-a} \mathrm{e}^{-\left(\frac{r}{\xi_{\rm L}}\right)^b} + (L-r)^{-a} \mathrm{e}^{-\left(\frac{L-r}{\xi_{\rm L}}\right)^b} \ .
\end{equation}
These fits also allow us to extract the exponents $a$ and $b$, which we find to have a strong dependence on $\varphi$, as shown in Fig.~\ref{fig:xi}c and d.
The oscillations of $G_\mathrm{P}(r)$ at low values of $r$, however, make impossible an accurate extraction of $a$ from fitting. For this reason, we impose $a'\sim 1$ (an intermediate value of $a$ in Fig.~\ref{fig:xi}c) and $b'=1$, and extract $\xi_{\rm P}$ from fitting $G_\mathrm{P}(r)$ to
\begin{equation}\label{eq:GP}
G_\mathrm{P}(r)\sim r^{-1}\mathrm{e}^{-\frac{r}{\xi_{\rm P}}}.
\end{equation}
The results for this second correlation length are shown in
Fig.~\ref{fig:xi}b.  Both estimators are expected to
  measure the same object, that is, both $\xi_{\rm L}$ and $\xi_{\rm
    P}$ should be proportional to the true correlation length, $\xi$.
  The actual values obtained from both fits must, however, be regarded merely
  as indicators of the correlation growth, because the extraction
  of $\xi$ is rather inaccurate. The linear size of 
  simulation box should be several times larger than the correlation length $\xi$ to
  obtain accurate estimations.

The growth in the heterogeneity of the system can be visualized by
looking at the spatial fluctuations of $\Delta_\mathrm{AB}$. Figure~\ref{fig:snapshots} 
shows the typical cages of all the
particles at different  $\varphi$ along the compression process. The
 cage of particle $i$ ($i=1,\ldots,N$) is represented as a
sphere of diameter $D_i=\alpha |\V{r}_i^A-\V{r}_i^B|$
centered in $\vr_i^A$. The normalization $\alpha$  fixes
the average cage size, $\frac{1}{N}\sum_{i}^N D_i=\overline{\sigma}=1$,
for the sake of visualization.

\section{Phase diagram for thermal glasses}

\begin{figure}[h]
\centerline{\hbox{ \includegraphics [width =0.7\columnwidth] {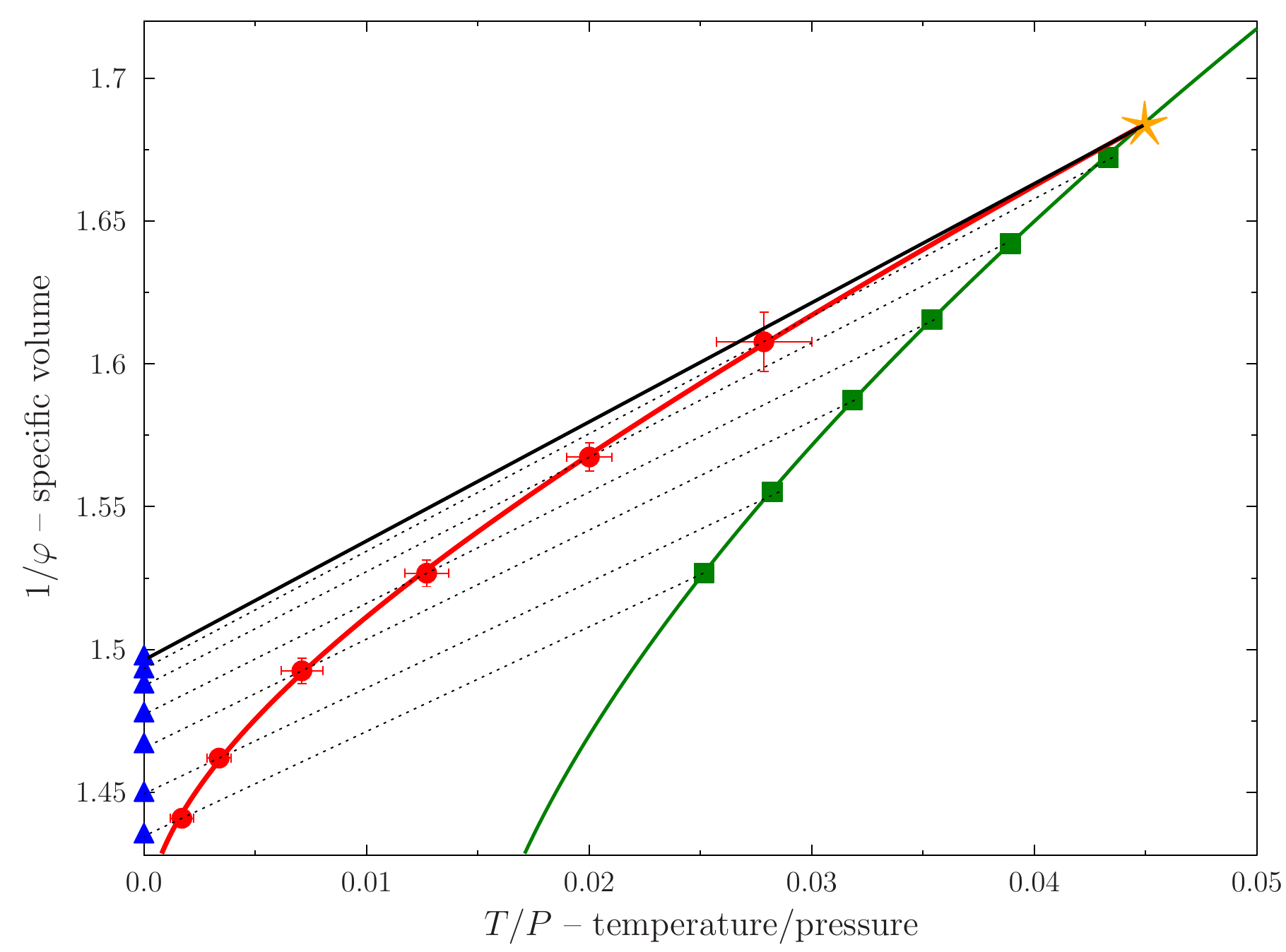} }}
\caption{Phase diagram for thermal glasses. The data from Fig. 1 (main text) are rescaled to $1/\varphi$ versus $T/P = 1/(\rho p)$. Supercooled liquid states are equilibrated at 
 the glass transition temperature $T_{\rm g}$ (green squares) below the  dynamical crossover temperature $T_{\rm d}$ (gold star), and are annealed (dashed lines) to their zero-temperature ground states (blue triangles). The stable glasses transform into marginally stable glasses at the Gardner temperature $T_{\rm G}$ (red circles and line).}
\label{fig:PD_TE}
\end{figure}

In order to connect  our $1/p-\varphi$ phase diagram for HS (Fig. 1 in the main text) with traditional presentations of thermal glass results (see, e.g, Fig. 1 in Ref.~\cite{DS01}), we present an alternate version of that phase diagram (Fig.~\ref{fig:PD_TE}). In this different representation, we plot the specific volume $1/\varphi$ as the $y$ axis and the ratio between temperature and pressure $T/P = 1/(\rho p)$ as the $x$ axis. It essentially describes how the specific volume changes with the temperature, at constant pressure, in different phases. We expect this phase diagram to be qualitatively reproducible in thermal glass experiments.

\section{Bidisperse hard disk results and analysis}

\begin{figure}[h]
\centerline{\hbox{ \includegraphics [width =0.7\columnwidth] {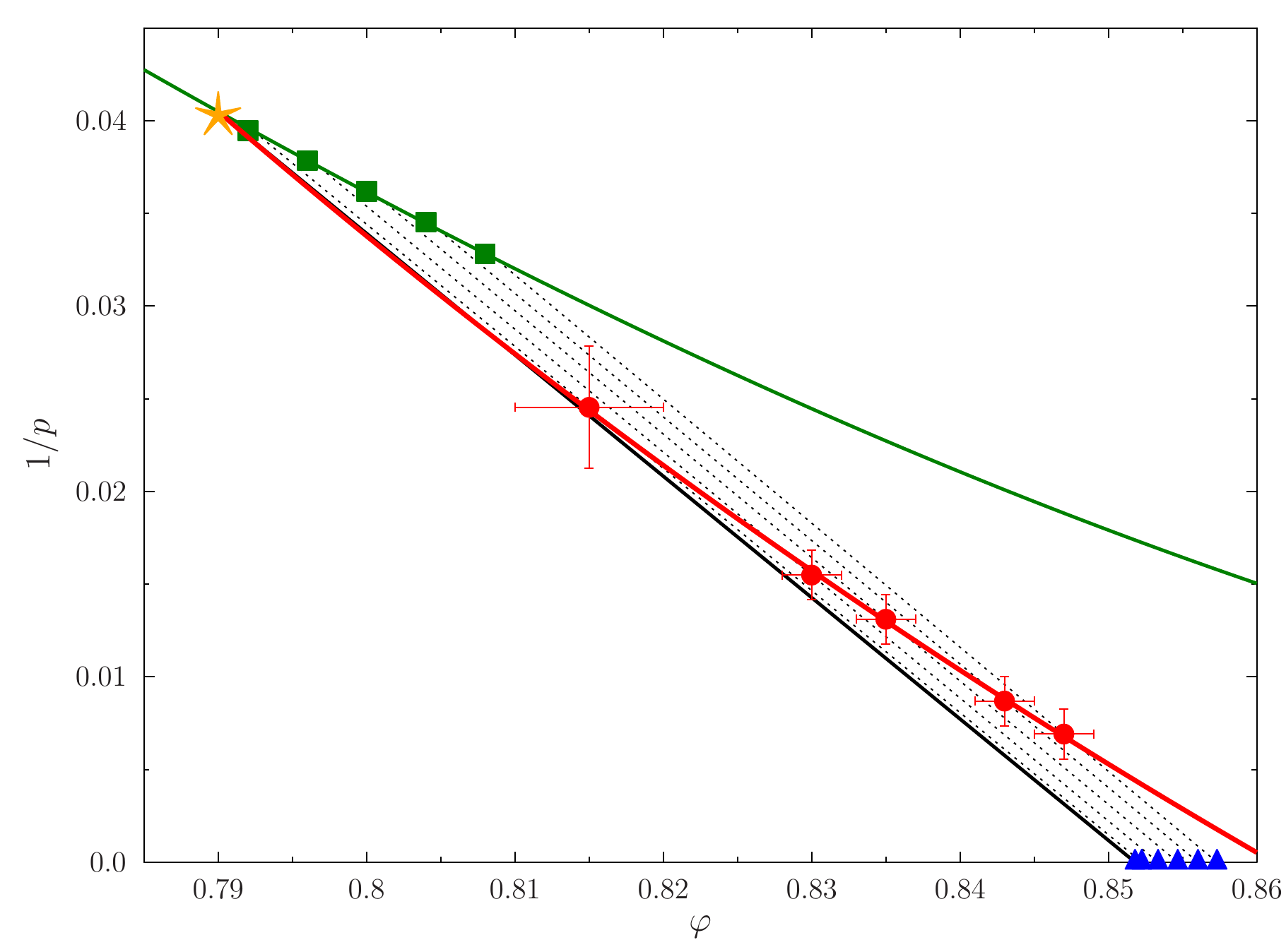} }}
\caption{Phase diagram for bidisperse HD. Symbols are the same as in Fig. 1 of the main text. The 2$d$ liquid EOS Eq.~(\ref{eq:2dEOS}) is used.}
\label{fig:PD2d}
\end{figure}

We also study a two-dimensional bidisperse model glass former~\cite{Berthier2014PRL}, using the same approach as for HS described the main text. 
The system consists of an equimolar binary mixture of $N=1000$ hard disks (HD) with  diameter ratio $\sigma_1:\sigma_2 = 1.4:1$.
In this case, we do not use the swap algorithm.
Equilibrium configurations are obtained by slow relaxations during MD runs, so that particles all diffuse, i.e.,  $\Delta(t) \geq 10 \sigma_1^2$. For each $\varphi_{\rm g}$, $N_{\rm s}=100$ samples are obtained. The liquid EOS is fitted to
\beq
p_{\rm liquid}^{2d}(\ph) = 1 + f^{2d}(\varphi)[p_{\rm CS}^{2d}(\varphi)  - 1],
\label{eq:2dEOS}
\eeq
where
\beq
p_{\rm CS}^{2d}(\varphi) = 1 + 2 \ph \frac{1-c_1 \ph}{(1-\ph)^2}
\eeq
is the 2$d$ Carnahan-Starling (CS) form, and 
\beq
f^{2d}(\varphi) = 1 + c_2 (1+c_3 \ph^{c_4})
\eeq
is a fitted function with parameters $c_1 = 0.52$, $c_2=1.0$, $c_3=2.7$, and $c_4=14$.
The estimated dynamical crossover
is $\phid=0.790(1)$.
The $d=2$ version of $\varphi_{\rm G}$ is obtained from the peak of caging skewness of big particles. The results are summarized in Table~\ref{tab:phiG2d} and the  phase diagram is reported in Fig.~\ref{fig:PD2d}.

\section{Summary of numerical results}
We summarize numerical values of our main results for HS in Table~\ref{tab:phiG3d} and for HD in Table~\ref{tab:phiG2d}.

\begin{table}[h]
\caption{Numerical values of $\phij$ and $\phiG$ for polydisperse HS.
}
\label{tab:phiG3d}
\begin{tabular}{ccc}
 $\phiin$&$\phij$&$\phiG$\\\hline
0.598 &0.670(1) &   0.622(4) \\
0.609&0.672(1) &  0.638(2)  \\
0.619&0.677(1) &0.655(2)  \\
0.630&0.682(1) &0.670(2)  \\
0.643&0.690(1)& 0.684(1) \\
0.655&0.697(1)&0.694(1)\\
\end{tabular}
\end{table}

\begin{table}[h]
\caption{Numerical values of $\phij$ and $\phiG$ for bidisperse HD.}
\label{tab:phiG2d}
\begin{tabular}{ccc}
 $\phiin$&$\phij$&$\phiG$\\\hline
 0.792&0.852(1)& 0.815(5)\\
 0.796&0.853(1)& 0.830(2)\\ 
 0.80&0.855(1)&0.835(2)\\
 0.804&0.856(1)&0.843(2)\\ 
 0.808 &0.857(1)&0.847(2)\\
\end{tabular}
\end{table}


\bibliographystyle{pnas}
\bibliography{HS,glass,Gardner}
\end{article}
\end{document}